\documentclass[preprint2]{aastex}

\newcommand{\myemail}{wnp@mpe.mpg.de}

\def\cm-2{cm$^{-2}$}

\def\HII{\hbox{H\,{\sc ii}}}

\def\ein{{\it Einstein}}
\def\asca{{ASCA}}
\def\chandra{{\it Chandra}}

\def\xmm{{XMM-Newton}}
\def\ro{{ROSAT}}

\def\cm33{ChASeM33}
\def\n253{\object{NGC~253}}
\def\me31{\object{M~31}}
\def\m33{\object{M~33}}
\def\mx7{\object{M~33~X$-$7}}
\def\xe7{X$-$7}
\def\x47{PMH47}
\def\xf47{[PMH2004]~47}

\newcommand{\ergcm}[1]{$\times 10^{#1}$ \hbox{erg cm$^{-2}$ s$^{-1}$}}

\newcommand{\ergs}[1]{$\times 10^{#1}$ \hbox{erg s$^{-1}$}}
\newcommand{\oergs}[1]{$10^{#1}$ erg s$^{-1}$}
\newcommand{\hcm}[1]{$\times 10^{#1}$ cm$^{-2}$}

\newcommand{\expo}[1]{$\times 10^{#1}$}

\newcommand{\nh}{\hbox{$N_{\rm H}$}}

\newcommand{\ct}{ct s$^{-1}$}

\newcommand{\MSOL}{\mbox{$\:M_{\sun}$}}

\shorttitle{The second eclipsing HMXB in M~33}
\shortauthors{Pietsch et al.}

\begin{document}


\title{Detection of the second eclipsing high mass X-ray
binary in M~33}


\author{Wolfgang Pietsch and Frank Haberl}
\affil{Max-Planck-Institut f\"ur extraterrestrische Physik, Giessenbachstrasse,
       85741 Garching, Germany}
\email{\myemail}

\author{Terrance J. Gaetz, Joel D. Hartman, Paul P. Plucinsky, Ralph T\"ullmann}
\affil{Harvard-Smithsonian Center for Astrophysics, 60 Garden Street, Cambridge,
MA 02138}

\and

\author{Benjamin F. Williams}
\affil{Astronomy Department, University of Washington, Box 351580, Seattle, WA
98195}

\and

\author{Avi Shporer and Tsevi Mazeh}
\affil{School of Physics and Astronomy, Raymond and Beverly Sackler Faculty of
Exact Sciences, Tel Aviv University, Tel Aviv 69978, Israel}

\and

\author{Thomas G. Pannuti}
\affil{Space Science Center, Morehead State University, 200A Chandler Place,
Morehead, KY 40351}

\begin{abstract}
\chandra\ data of the X-ray source \xf47\  were obtained in the 
ACIS Survey of \m33\ (\cm33) in 2006.  During one of the 
observations, the source varied from a high state to a low state and 
back, in two other observations it varied from a low 
state to respectively intermediate states. These transitions are interpreted 
as eclipse ingress and egresses of a
compact object in a high mass X-ray binary system. 
The phase of mid eclipse is given by  
HJD~245\,3997.476$\pm$0.006, the eclipse half angle is $30.6\degr\pm1.2\degr$. 
Adding \xmm\ observations of \xf47\ in 
2001 we determine the binary period to be 1.732479$\pm$0.000027~d. This period is
also consistent with \ro\ HRI observations of the source in 1994.
No short term periodicity compatible with a rotation period of the compact
object is detected. There are indications for a long term variability similar 
to that detected for Her X--1.
During the high state the spectrum of the source is hard (power
law spectrum with photon index $\sim$0.85) with an unabsorbed luminosity 
of 2\ergs{37} (0.2--4.5 keV). 
We identify as an optical counterpart a $V \sim 21.0~{\rm mag}$ star with
$T_{\rm eff} > 19000~{\rm K}$, $\log(g) > 2.5$. CFHT optical light curves for
this star show an ellipsoidal variation with the same period as the X-ray
light curve.
The optical light curve together with the X-ray eclipse
can be modeled by a compact object with a mass consistent with a neutron star 
or a black hole in a high mass X-ray binary. However, the hard power law X-ray
spectrum favors a neutron star as the compact object in this second eclipsing
X-ray binary in \m33. Assuming a neutron star with a canonical mass of 
1.4 \mbox{$\:M_{\sun}$} and the best fit companion temperature 
of 33000~K, a system inclination $i = 72\degr$ and a 
companion mass of 10.9 \mbox{$\:M_{\sun}$} are implied. 

\end{abstract}

\keywords{Galaxies: individual: \m33\ --- X-rays: individuals: 
[PMH2004] 47 --- X-rays: binaries --- binaries: eclipsing}

\section{Introduction}
Only a few eclipsing X-ray binaries (XRBs) have been detected in the Milky Way and other
nearby galaxies. However, eclipsing XRBs are very interesting sources as they
provide (when X-ray and optical data are included) the most reliable detailed mass
measurements for compact stellar remnants possible \citep[with the exception
of the few known radio pulsars in binaries, including double neutron stars, see
e.g.][]{2006csxs.book....1P}.

Of specific interest is the first eclipsing XRB in \m33, \mx7\ 
(hereafter \xe7), which showed properties that were not known from XRBs in the
Galaxy or the Magellanic Clouds. 
\xe7\ was already 
detected as a variable by the \ein\
observatory with a luminosity at maximum in excess of \oergs{38} 
\citep[][]{1981ApJ...246L..61L,1983ApJ...275..571M} 
and it stayed active in all following observations. 
Its variability was explained through the model of 
an eclipsing XRB with an orbital period of 3.45~d and an eclipse duration
of $\sim$0.4~d  
based on \ro\ and \asca\ data
\citep[][]{1997AJ....113..618L,1999MNRAS.302..731D}. With the help of the 
well-constrained \chandra\ position of \xe7, 
\citet[][]{2004A&A...413..879P}  identified a B0I to O7I star of 18.89 mag in 
$V$ as optical counterpart: this star
shows the ellipsoidal light curve of a high mass XRB (HMXB)
with the \xe7\ binary period. 
They argued that the compact
object in the system is a black hole based on the mass of
the compact object derived from orbital parameters and the optical 
companion mass, the lack of pulsations, and analysis of the extracted 
X-ray spectrum. 
Those authors concluded
that \xe7\ would be the first eclipsing black hole HMXB.

The black hole nature of \xe7\ was firmly established by 
\citet[][]{2006ApJ...646..420P} using observations of the 
\chandra\ Advanced CCD Imaging Spectrometer (ACIS) survey of \m33\ \citep[\cm33, see][]{2008ApJS..174..366P}
which accumulated in seven ACIS-I pointing directions 
a total exposure of 200~ks each. During several of these pointings
\xe7\ was in the field of view. The \cm33\ measurements of 
\xe7\ resolved for the first time the eclipse ingress and egress 
as well as constrained the light curve of \xe7\ for binary phases around 
eclipse. 
In addition, \citet[][]{2006ApJ...646..420P} identified \xe7\  on archival HST WFPC2
images. Through detailed modeling of the optical light curve, the observed 
X-ray eclipse and 
improved parameters for the companion star, a lower mass limit for the
compact object in the system of 9~\MSOL\ \citep[see also][]{2007A&A...462.1091S}
has been obtained.
The black hole mass was further
constrained as 15.65~\MSOL\ with the help of 
a radial velocity curve based on Gemini North spectra and careful
modeling \citep[][]{2007Natur.449..872O}. This -- at the time -- made \xe7\ the 
stellar black hole with the highest mass that had been determined with high
precision. 

In this paper we report on the detection and optical identification of the
second eclipsing XRB in \m33\ within the \cm33\ project. The source
was already detected in the \xmm\ survey of \m33\ of 
\citet[][hereafter PMH2004]{2004A&A...426...11P} and is source  no. 47 in their
catalogue (\xf47, hereafter \x47).
During the \xmm\ observations it showed strong time
variability and was classified as a transient XRB candidate by
\citet[][]{2006A&A...448.1247M}. 
It was first reported in the \m33\ 
\ro\ catalogue by \citet[][]{2001A&A...373..438H} as source 19. 
\x47\ is listed as  no. 8 in the ``first look" \cm33\ catalogue 
\citep[][]{2008ApJS..174..366P}.
In several \cm33\ observations the source was
detected far off axis with strong time variability which 
-- in a preliminary analysis -- could be 
explained as the signature of a 1.7 d eclipsing XRB 
\citep[][]{2006ATel..905....1P}. 
A massive star from the Local Group survey catalogue of \m33\ stars
\citep[][]{2006AJ....131.2478M} could be identified as the optical counterpart of the
X-ray source. In a re-analysis of the deep 3.6m Canada-France-Hawaii Telescope 
(CFHT) photometric survey images of \m33\ 
\citep[][]{2006MNRAS.371.1405H}, the proposed optical companion  
showed the ellipsoidal light curve of
a HMXB in \m33\ \citep[][]{2006ATel..913....1S}. 
A coherent analysis of the \chandra\ and \xmm\ X-ray and optical data allows us
to further constrain the system parameters. Our follow-up analysis of \x47\ 
in the \ro\ data also indicates time variability consistent with the 
\chandra\ and \xmm\ ephemeris. 

\section{Observations and data analysis}\label{sec:obs}
The \x47\ area was covered by the ACIS CCDs during seven observations of 
the \cm33\ project and during the archival observation with observation
identification (ObsID) 1730.
Table~\ref{tbl:obs_chandra} summarizes these observations
giving the \cm33\ field in column 1, ObsID 
(2), observation start date (3), 
sum of good time intervals (``on-time", 
4), the ACIS chip covering \x47\ (5), the offset of the source from the pointing 
direction (6), and the \x47\ binary phase and cycle number during the 
observation (7, 8) using ephemeris that will be
discussed in Sect.~\ref{sec:eph}. In the last column (9), we comment on source
brightness and observed binary features (see Sect.~\ref{sec:tim} and
\ref{sec:eph}).   
The average brightness of \x47\ varied strongly between the \cm33\  
observations. 
However, the determination of accurate count rates and light curves turned out 
to be quite difficult because in all observations the source is positioned far 
off axis and therefore has a large extent
due to broadening of the telescope point spread function (PSF) which may stretch over bad CCD 
columns (e.g. in ObsID 6377)
or may partly fall outside the detector field of view (FoV)
(6382, 6385, 6387, 7226, 7344). An extreme case is presented in ObsID 6383 where
the position of the source falls outside the detector and only a small part of
the PSF is recorded. In several of the observations,
counts from \x47\ are rejected in standard level 2 event files when the source
is moving across rejected columns due to satellite dithering. This effect can
reduce the counts in 1000 s integration intervals by varying amounts and create
spurious periods in timing analysis. Therefore, we created new level 2 event
files for broadband time variability analysis that did not reject these columns. 
For better comparison we applied an approximate flux correction to account for 
the size of the extraction region relative to the PSF size and to
convert to an equivalent on-axis source.
For each observation, a  
{\tt ChaRT}\footnote{{\tt http://cxc.harvard.edu/chart/}}
raytrace \citep{2003ASPC..295..477C} 
was performed for a 1.5 keV on-axis point source for an input
ray density of 1 mm$^{-2}$, and the total number of resulting rays was 
evaluated.  For each observation, another {\tt ChaRT} run was performed for a 
1.5 keV point source (ray density also 1 mm$^{-2}$)
at the appropriate off-axis position; this applies the
mirror reflectivity and vignetting.  The rays were projected to
the detector using the CIAO tool {\tt psf\_project\_ray}, which also 
applied an approximate detector quantum efficiency (QE).  
In many cases, the source is near
the edge of the detector, or in one case, off the detector (although
some of the large PSF at that position is still on the detector).
To account for these exposure variations resulting from chip edges
and spacecraft dither, we weighted ``counts'' images generated from
the raytrace pseudo-event lists by a fractional exposure map.  The
fractional exposure map accounts for the fraction of the time a
sky pixel is on the detector, and is obtained by constructing
an ``exposure map'' in which the \chandra\ High Resolution Mirror Assembly 
(HRMA) effective area and 
detector QE are both set to unity.  We extracted weighted counts
from a region corresponding to the extraction region in the real
observation to get the predicted number of counts
within the extraction region used for the real data.  This
provides the predicted number of counts for the given source intensity.
The normalization to an on-axis source is obtained by dividing the
on-axis predicted counts by the extracted off-axis predicted counts.
This factor was used to scale the observed fluxes to the values
they would have had on-axis.
For spectral fitting, \x47\ photons
were extracted from CIAO v4.0 level 2 event files using the same area as for 
the time variability analysis, but with the bad columns excluded. 
Spectra were grouped 
in channels with at least 20 photons in the on-source spectra.

In the \xmm\ EPIC \m33\ raster project, the \x47\ area was covered
in seven observations 
at different off-axis angles (typical ontime of 14 ks)
spanning about 
three years.
Table~\ref{tbl:obs_xmm} summarizes these observations
giving observation identification (ObsID) in column 1, 
observation start date (2), on-time (3),
the offset of the source from the 
pointing direction (4),  0.2--4.5 keV luminosity 
\citep[Col. 5, derived from
Table 5 of ][]{2006A&A...448.1247M},  
and the \x47\ binary phase and cycle number during the 
observation (6, 7) using ephemeris that will be
discussed in Sect.~\ref{sec:eph}. In the last column (8), we comment on source
brightness and observed binary features. 
The X-ray source was detected only during two observations in July 2001 
which were separated by about 2.5 days.
However, during an observation between these two, the
source was not detected.
For the \xmm\ EPIC X-ray spectral analysis we used PN (single +
double pixel events, PATTERN 0--4) and MOS (PATTERN 0--12) events 
disregarding bad CCD
pixels and columns (FLAG 0). The EPIC spectra were simultaneously fit for the
three instruments 
allowing for a constant normalization factor between the spectra. Spectra were 
grouped in channels with at least 30 photons in the on-source spectra.

In the \ro\ catalogue of \m33\ \citep[][]{2001A&A...373..438H}, the detection of 
\x47\ was reported in the HRI at a level of significance greater than 7$\sigma$ 
while in the PSPC it is only marginally detected at a level of significance
of only
3$\sigma$. We reanalyzed the \ro\ HRI data of ObsIDs 600487h, 600488h and
600489h performed from July 27 to August 8, 1994, where \x47\ is in the FoV. 
After screening for high background, we combined data with continuous
observation intervals that led to variable integration times of typically 1700~s
(minimum 786~s, maximum 2514~s), depending on the duration of the scheduled
observation and background. 

The \x47\ field was covered by the CFHT variability survey of \m33\ conducted in
the  $g'$, $r'$, and $i'$ bands during 27 nights from 2003 August to 2005 January. 
For details of this survey and the data reduction procedure used see 
\citet[][]{2006MNRAS.371.1405H}.   
To summarize the procedure, the
standard CCD calibrations were applied to the images as part of the
CFHT Queue Service Observing mode. We performed image subtraction
photometry \citep[][]{1998ApJ...503..325A,2000A&AS..144..363A}
on these images to
obtain differential flux light curves for the source and PSF fitting
photometry on a stacked reference image using the DAOPHOT/ALLSTAR
package \citep[][]{1987PASP...99..191S,1992ASPC...25..297S}
to obtain the reference flux values. We
then applied a scaling factor to the formal photometric uncertainties
determined by requiring that $\chi^{2}$ per degree of freedom
scatters about unity for $\sim 1000$ non-variable stars spanning the
full dynamic range of magnitudes.

For the data analysis we in addition used tools in the ESO-MIDAS v05SEPpl1.0, 
EXSAS v03OCT\_EXP, CIAO v3.2 and v4.0, and HEAsoft v6.3
software packages as well as the imaging application DS9 v4.13 and the software
package {\tt ACIS Extract} \citep[][]{Broos2002}.

\section{Improved position}
\begin{figure*}
\includegraphics[height=10cm,clip]{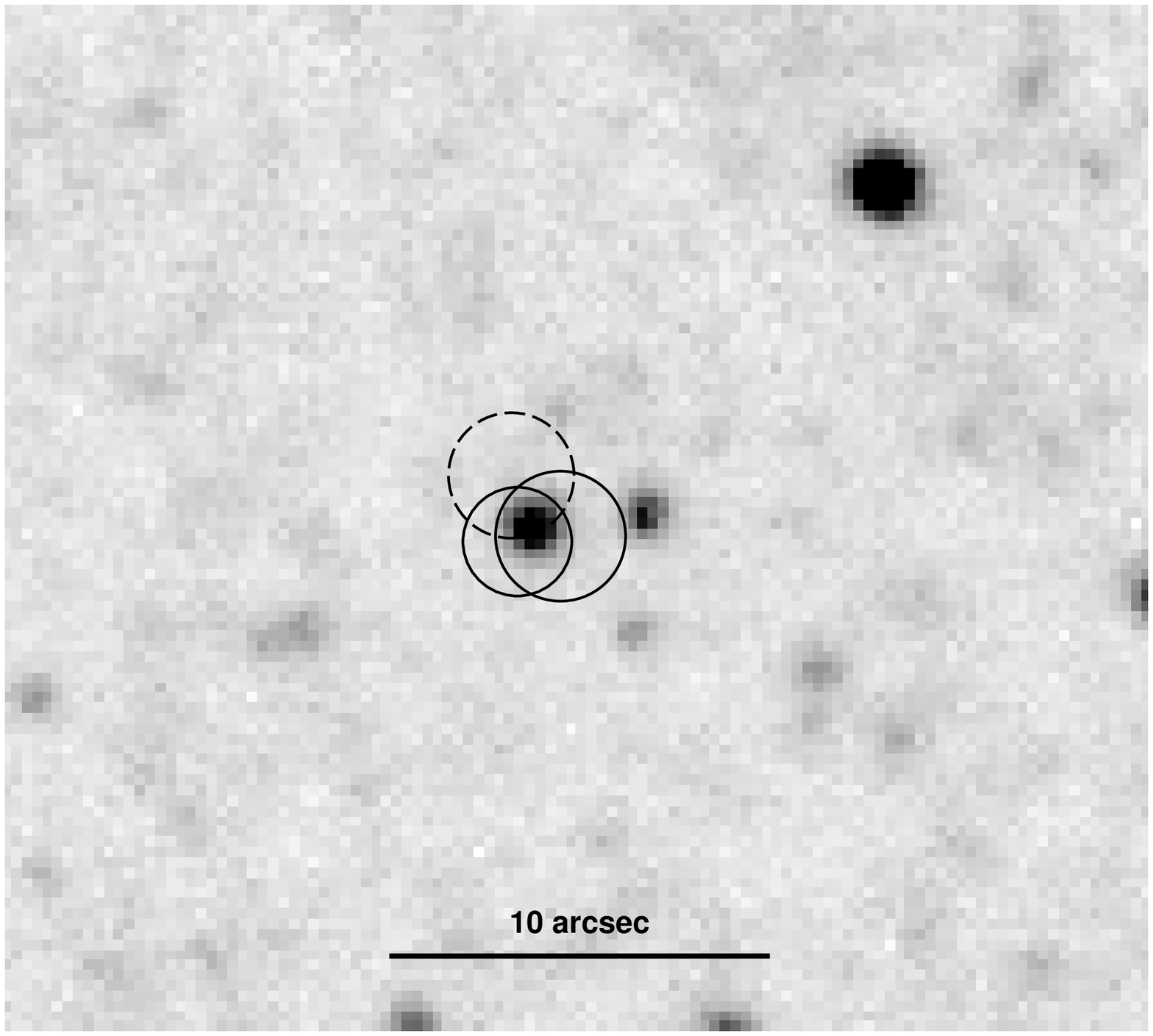}
\caption{Local Group survey B image \citep[][]{2006AJ....131.2478M} of the
\xf47\ field.
X-ray positions are indicated by circles with radii of the total 
3$\sigma$ error. 
The circle to the SW gives the
\xmm\ position 
\citep[][]{2006A&A...448.1247M}. 
The dashed circle to the NE indicates the \cm33\ position 
\citep[][]{2008ApJS..174..366P}.
The small circle indicates the improved \cm33\ position (see text).
The bright object included in all error circles is LGGS J013236.92+303228.8
 \citep[][]{2006AJ....131.2478M}. 
\label{fig:fc}}
\end{figure*}

The best \xmm\ position for \x47\ was derived as 
RA$_\mathrm{J2000}$ = 01$^{h}$32$^{m}$36\fs85, 
Dec$_\mathrm{J2000}$ = +30\degr32\arcmin28\farcs5
with a 1$\sigma$\ error of 0.57\arcsec\ including systematics 
\citep[][]{2006A&A...448.1247M}.
In the \cm33\ ``first look" paper \citep[][]{2008ApJS..174..366P} the position of \x47\ is given as 
RA$_\mathrm{J2000}$ = 01$^{h}$32$^{m}$36\fs95, 
Dec$_\mathrm{J2000}$ = +30\degr32\arcmin30\farcs1
with a 1$\sigma$\ statistical error of 0.48\arcsec.
Assuming for \x47\ the same systematic error 
as for sources within 3\arcmin\ from the aim point\footnote{see
http://cxc.harvard.edu/cal/ASPECT/celmon/} a 99\% error of 0.8\arcsec\ has to be
added.
We obtained the most reliable source position for ObsID 6385, as the 
vast majority of the PSF was contained on the ACIS-I array in this 
observation and the source was bright during most of the integration. 
Our {\tt ACIS Extract} analysis of this observation determined the mean 
position of the source counts within the 90\% encircled energy PSF for 
the sources in the image, as well as the source position that provided 
the best match to a cross-correlation between the distribution of source 
counts and the PSF shape at this location on the ACIS-I array.  These 
two position determinations were in very good agreement for \xe7\ and 
\x47, both of which were contained in this observation.  We then 
determined the offset between the \xe7\ position and its known optical 
counterpart in the \citet[][]{2006AJ....131.2478M} catalogue.  
This offset we assume 
to be the systematic offset between the \chandra\ X-ray image and the 
\citet[][]{2006AJ....131.2478M} catalogue images.  The offset was +0.5\arcsec\ in RA and 
+0.3\arcsec\ in Dec, with an alignment error of $\pm$0.3\arcsec.  After making 
this correction, the final position and error determined for \x47\ in 
ObsID 6385 was 
RA$_\mathrm{J2000}$ = 01$^{h}$32$^{m}$36\fs94,
Dec$_\mathrm{J2000}$ = +30\degr32\arcmin28\farcs4
with a 1$\sigma$\ statistical error of 
$\pm$0.37\arcsec\ corresponding to a 1$\sigma$\ total error of 0.48\arcsec. 
In Fig.~\ref{fig:fc} we show the improved position
together with the \xmm\ position \citep[][]{2006A&A...448.1247M} and that from 
the \cm33\ ``first look" paper overlaid on an optical image of the Local Group
Survey \citep[][]{2006AJ....131.2478M}.

\section{X-ray time variability}\label{sec:tim}
For the \cm33\ data, we sampled background and solar system barycentre corrected 
light curves of \x47\  with a time resolution from 1000~s to 5000~s depending on
the brightness of the source in the observation. To increase the signal 
to noise in the far off-axis observations, we
restricted the analysis to the 0.5--5~keV band which covers most of the source
flux (see Fig.~\ref{fig:lc_chandra}). In addition we indicated in the images to 
the right of the light curves the extraction regions for source and
background. Rectangular regions were used in some cases to get similar
background structure in source and background regions.   
The dashed ellipses show the HRMA PSF 90\% enclosed counts fraction. 
For observations far off axis, we used extraction areas smaller than  90\% PSF 
to improve the signal to noise ratio (e.g. for ObsIDs 1730).    
The light curve information has been 
confirmed by investigating time-selected images.
In the following we give a short description of the intensity behavior of the
source within the individual observations. 

During ObsIDs 1730 and 7344 no flux is
detected from the source (3$\sigma$ ACIS-I on-axis upper limit of 
8\expo{-3} \ct\ in the 0.5--5 keV band in 5000 s integration intervals).
 
During ObsID 6382, the source is faint and we needed to integrate the data
over 5000~s. During the first 10 ks of the observation (untill 
HJD~2453698.164$\pm$0.029) no significant flux is 
detected.

During the short ObsID 7226, the source is clearly detected but faint. The 
light curve again had to be integrated over 5000~s. No significant variability is 
detected. 

During ObsID 6383 flux at the wing of the PSF of the source is clearly
seen in the image. However, from HJD~2453902.052$\pm$0.029 to 
HJD~2453902.342$\pm$0.029 no photons at all are detected at that position.

During ObsID 6387, \x47\ is at least partly detected in a high
state which allowed us to sample the light curve with 1000~s resolution. 
We observed a transition of the source from zero 
to high intensity at HJD~2453912.740$\pm$0.012 about 15 ks after the start of
the observation. 

\begin{figure*}
\includegraphics[width=5cm,clip]{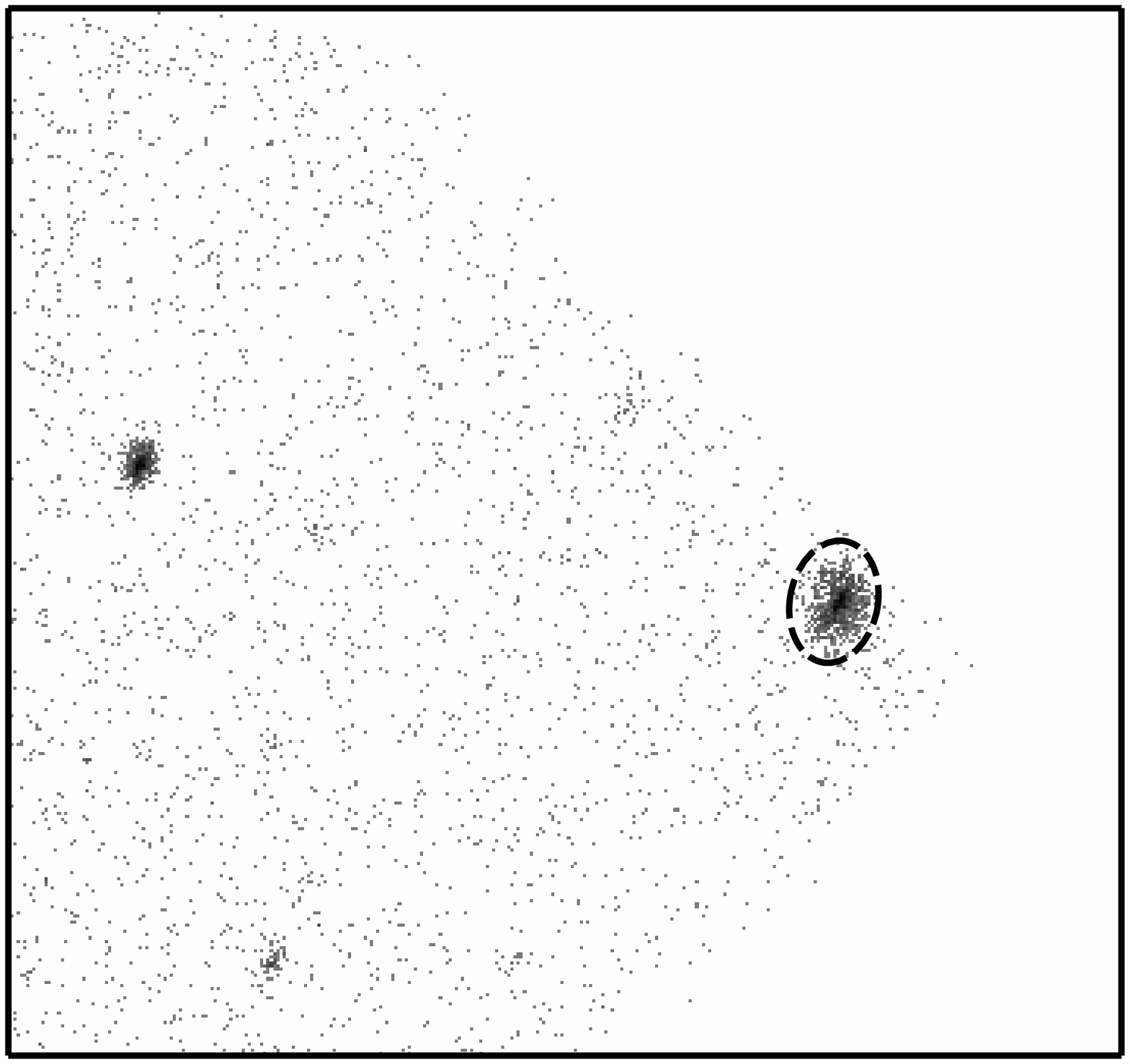}
\includegraphics[width=5cm,clip]{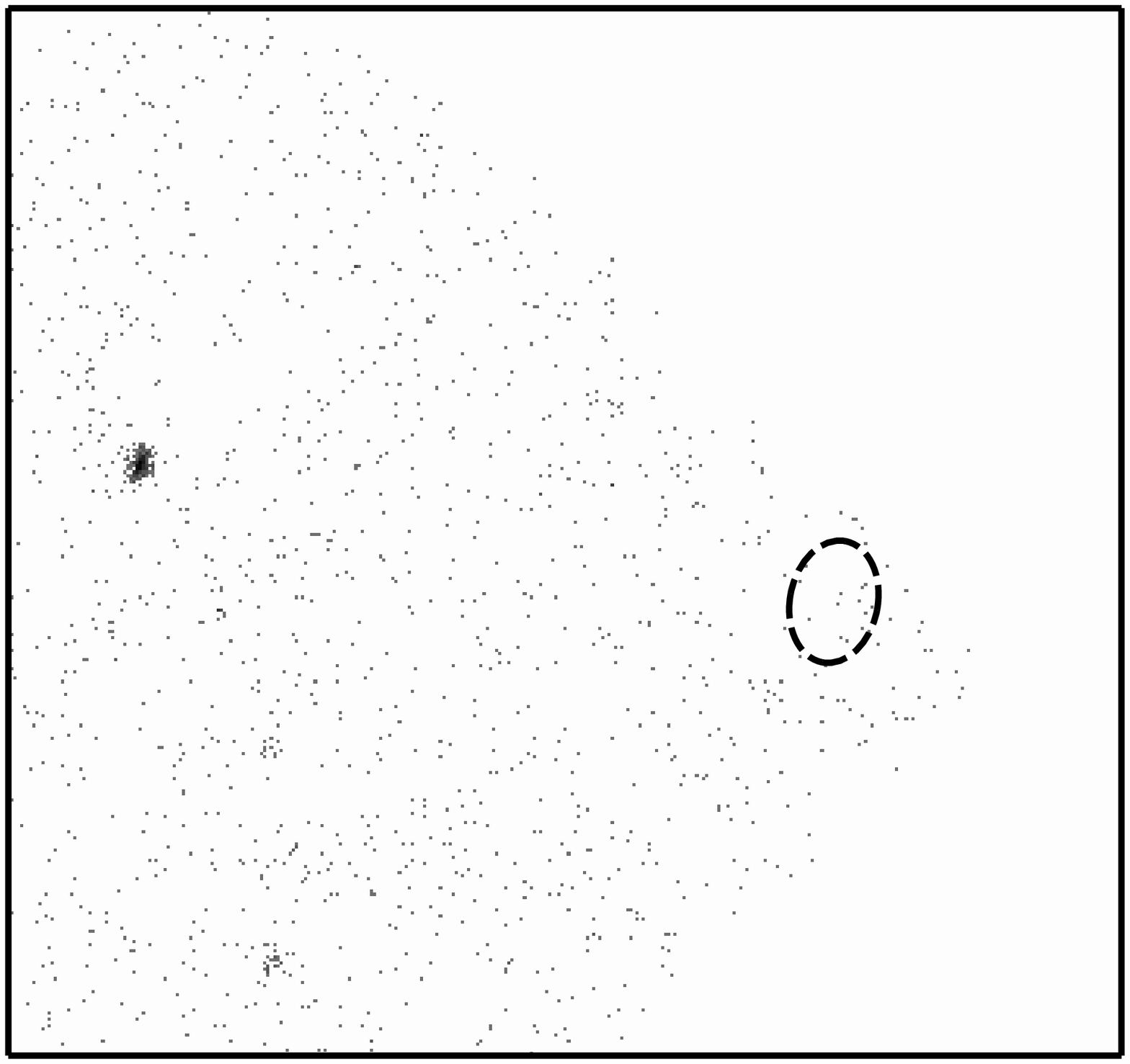}
\includegraphics[width=5cm,clip]{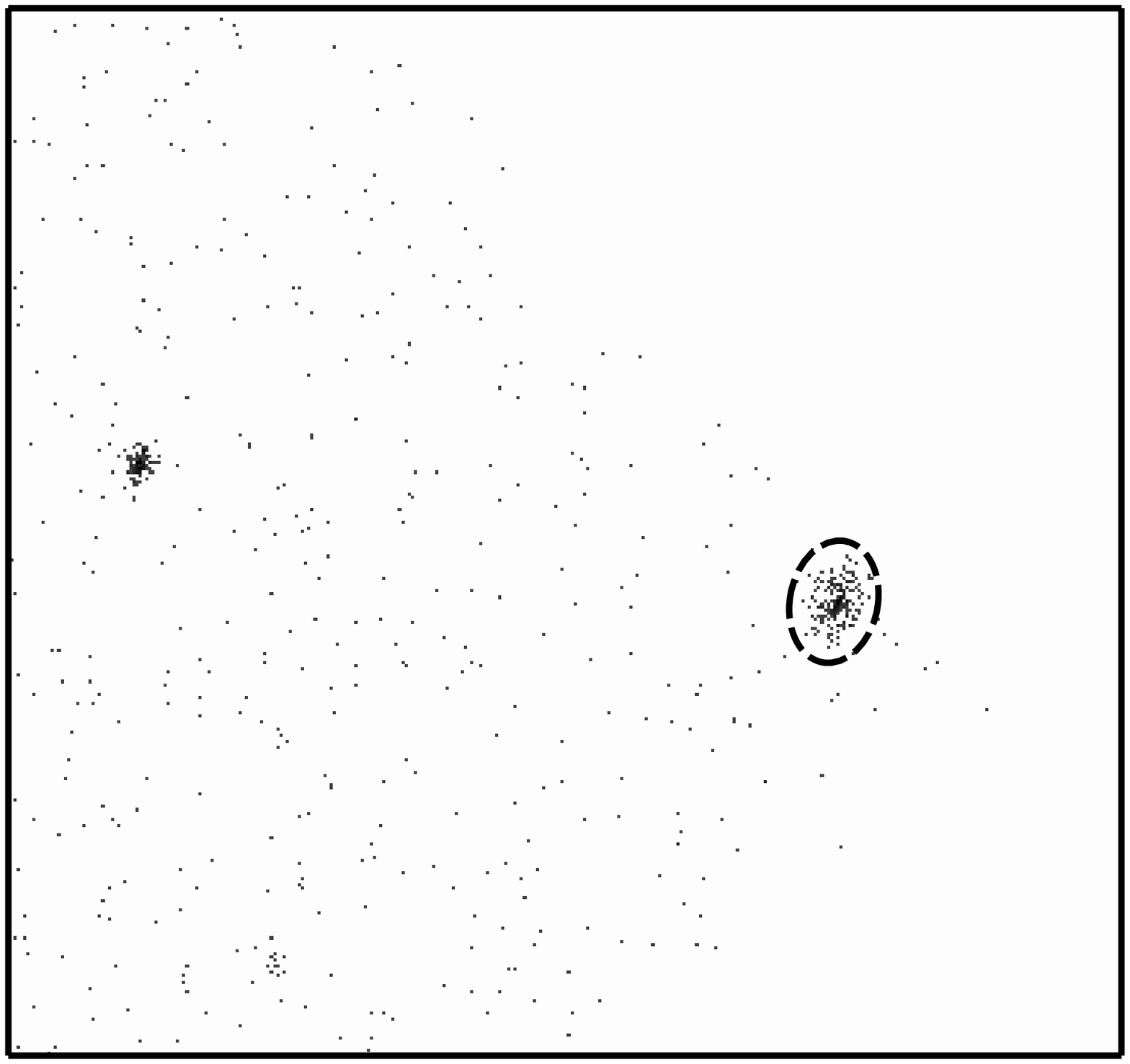}
\caption{$6\arcmin\times6\arcmin$ \chandra\ ACIS-I 0.5--5 keV images of observation 6385 for times before
the ingress of \xf47\ into low state (left), during low state (middle), and after
low state (right). The position of \xf47\ is indicated by the ellipse. 
\label{fig:ima_6385}}
\end{figure*}

During ObsID 6385 \x47\ is again for most of the time detected at high intensity
suitable for 1000~s binning.
We observed transitions of \x47\ from high intensity to
zero flux and back.  The source transits to the low state at 
HJD~2453997.329$\pm$0.006 and returns to the high state at 
HJD~2453997.624$\pm$0.006. Ingress into and egress out of the low state can 
not be resolved by the 1000~s time bins imposed by source statistics. The
duration of the low state is (25500$\pm$1000)~s. After the low state the source
seems to be more variable than before. Figure~\ref{fig:ima_6385}
demonstrates that during the low state no counts are detected from \x47.

During ObsID 6377, the source is not detected in the first $\sim$25 ks 
(until HJD~2454004.555$\pm$0.020) and then stayed in a faint state
with indications for variability. Due to the faintness of the source the light
curve was integrated with 3000~s resolution.

In the bright phase of \chandra\ ObsIDs 6385 and 6387 and in \xmm\ ObsID 0102641101 
we searched for pulsations in the
frequency range $10^{-4}$ Hz to 7 Hz and found no significant periodic
signal. 
For the \chandra\ observations, the 2$\sigma$ upper limit for a pulsation
amplitude is 25\% for frequencies below 0.15 Hz while for the \xmm\ observation, the
EPIC PN upper limit is 43\%  for frequencies below 7 Hz.  
As one expects a modulation of the photon arrival time with the orbital
phase which would smear out the pulsation signal for short periods \citep[amplitude of
less than 50 s for HMXBs with short orbital period, c.f.][]{1997ApJS..113..367B}, we not
only investigated power spectra for the entire observation, but also added up
power spectra for intervals as short as 3319 s (1024 time bins with the
\chandra\ instrument resolution of 3.241 s).

\section{Energy spectra}\label{sec:spec}
For the brighter states of \x47 -- that is, during the high states of \chandra\
ObsIDs 6385 and 6387 as well as \xmm\ ObsID 0102641101 -- more than 900 counts were
collected from the source, allowing us to perform a detailed spectral analysis. 
The resulting integration
times, raw count rates and degrees of freedom are listed in 
Table~\ref{tbl:spectra} together with the results of the power law fits.
For \xmm, luminosity refers to the EPIC PN spectrum, MOS2 yields
similar values while those for MOS1 are $\sim$7\% higher.

Due to the limited number of
photons we only fitted absorbed one component spectral models. We used two
absorption components, accounting for the Galactic foreground absorption 
\citep[with a
fixed hydrogen column density of 6\hcm{20} and elemental abundances
from][]{2000ApJ...542..914W} and the \m33\ absorption (with column density as a
free parameter in the fit and Galactic metal abundances). 
As expected for an X-ray binary, thin thermal plasma spectra with abundances 
fixed to solar do not give acceptable fits. Also thermal bremsstrahlung fits can
be rejected as the best fit temperature in all fits is only constrained by the
upper boundary (200~keV) mimicking a hard power law. A similar argument holds for
disk black body models. There, the best inner disk temperature values $T_{\rm
in}$ are around 5 keV
and the luminosity a few times \oergs{37}. According to 
\citet[][]{2000ApJ...535..632M}, such temperatures are not reached even in
ultra-luminous compact X-ray sources. Stellar mass black hole systems of similar
luminosity to \x47\ show inner disk temperatures well below 1 keV.
Power law fits for all observations have a similar or slightly worse minimum 
reduced $\chi^2$ compared to the disk black body models. They are consistent 
with a common photon index of 0.85. 
The \nh\ values for the power law model fits indicate some absorption 
within \m33\ or intrinsic to the source in addition to
the Galactic value \citep[5.97 and 6.32 $\times 10^{20}$ H cm$^{-2}$ in the 
direction
of \x47\ according to][respectively]{1990ARA&A..28..215D,1992ApJS...79...77S}.
Figure~\ref{fig:spec_6387} shows the spectrum of the bright state of \chandra\ 
ObsID 6387 as an example.
Un-absorbed source fluxes in the 0.2--4.5 keV band are in the range 
(2.4--2.7)\ergcm{-13} 
based on the best fitting power law model. These fluxes correspond to
source luminosities of (1.8--2.0)\ergs{37}, 
respectively. Throughout the paper, we assume a distance to \m33\ of 795~kpc 
\citep{1991PASP..103..609V}. 

The \nh\ values of the power law model fits indicates that \x47\ lies in the
plane of the \m33\ disk or even on the near side,
as the absorbing column within \m33\ can be determined 
to 1.4--1.9\hcm{21} from a
$47\arcsec\times93\arcsec$\ half power beam width H{\sc i} map 
\citep{1980MNRAS.190..689N}. Assuming total \nh\ values from 0.6--1.5\hcm{21} 
we can 
compute the expected optical extinction $A_{\rm V}$ to 0.34--0.84 mag 
and $E(B-V)$ to 0.12--0.28 using standard relations
\citep{1995A&A...293..889P}.

\section{Orbital period determination}\label{sec:eph}
The intensity changes of \x47\ from high to low and vice versa in the \chandra\ 
ObsIDs 6385, 6387 and 6377 are not resolved by the time resolution of our plots,
of 1000~s, 1000~s, and 
3000~s, respectively. Shorter time bins do not help as the
source is not bright enough. In the following discussion we try to interpret the 
intensity transitions reported in Sect.~\ref{sec:tim} by  eclipses 
of a compact X-ray source by a companion star in a binary system and to
determine eclipse parameters and orbital period of the system. For this
analysis we use ObsIDs 6385, 6387 and 6377 and compare the results to the count
rate behavior of \x47\ in the other observations.

From observation 6385 we derive the eclipse duration as (25500$\pm$1000)~s and a well
defined epoch of mid eclipse (HJD~245\,3997.476$\pm$0.006). However, the orbital period is
too long to detect a second eclipse during this 91 ks observation.

A minimum length for the binary period can be estimated as 1.04 days by adding up 
the eclipse duration in ObsID 6385 and the longest continuous time out of
eclipse (65000~s in ObsID 6387). To further constrain the
orbital period of the system we use times of eclipse egresses. 
In HMXB systems the time of eclipse egress in most cases
is better determined than the time of ingress which can be masked by additional
absorption due to the viewing
geometry through the innermost regions of the wind of the companion and dense
material following the compact object in its orbit 
\citep[e.g.][]{1992A&A...263..241H}. The two bright $>$90 ks ObsIDs 6385, 6387 
show an eclipse egress. Also ObsID 6377 and 6383 seem to indicate eclipses of
\x47. The frequency of the detection of eclipse in- and egresses in the 70 to
100 ks \chandra\ observations points at an orbital period of the
system that is not much longer than these observations. 

If we interpret the three transitions from low to high respective intermediate
state in ObsIDs 6385, 6387, and 6377 as eclipse egresses, 
the orbital period of the system can be determined. We use the fact that
the times of egresses have to be separated by an integer number of orbital
periods. We start from the shortest separation (ObsIDs 6385 and 6377, separation
$T_{\rm diff1} = 6.931\pm0.026$ d). Possible orbital periods are 
$P_{\rm n} = T_{\rm diff1}$/n
with n less than 7 (due to the shortest allowed orbital period of 1.04 days,
see above). We now can test
for which of these six candidate periods the time difference between the well
determined eclipse egresses of ObsIDs 6387 and 6385 
($T_{\rm diff2} = 84.884\pm0.018$ d) is consistent with an integer. Fortunately,
we find a unique solution.  For n = 4, $T_{\rm diff2}$  is -- within the errors
-- 49 times  $P_{\rm n}$ and we get a best period from the \cm33\ observations
of $P_{\rm \cm33} = 1.73233\pm0.00037$ d. 

This period can be further constrained by including \xmm\ observations. 
Extrapolating the eclipse egress from \chandra\ ObsIDs 6387 with
$P_{\rm \cm33}$ to the time of \xmm\ ObsIDs 0102641001 and 0102641101 
(1096 periods earlier) shows that an eclipse egress should have happened 
between these observations (separated by 0.046 d) as the source is not detected
until the end of ObsID 0102641001, but it is 
already bright at the beginning of ObsID 
0102641101. Combining \xmm\ and \chandra\ eclipse egress epochs we
obtain an improved orbital period for \x47\ of
$P = 1.732479\pm0.000027$ d. The eclipse duration corresponds to $0.170\pm0.007$
in phase or an eclipse half angle of $30.6\pm1.2$ degree. 

In Tables \ref{tbl:obs_chandra} and \ref{tbl:obs_xmm}  we give for the
individual observations binary phase and cycle number at the beginning and end 
of the
observations based on the best ephemeris determined above. It is clear that not
all the variability can be explained by eclipses and additional variability has
to be present. On the other hand some of the short \xmm\ observations entirely 
fall in eclipse. Also the time without photons in \chandra\ ObsID 6383 
(see Sect.~\ref{sec:tim}) coincides with an extrapolated eclipse.
 
\begin{figure*}
\includegraphics[width=10cm,clip]{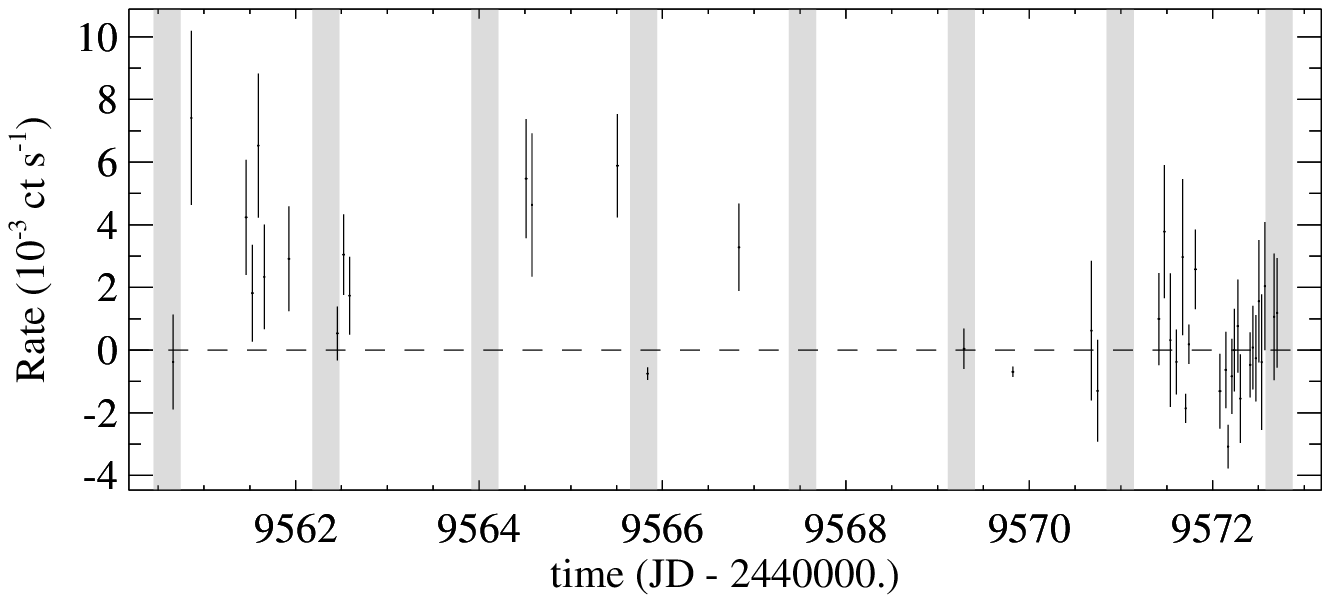}
\caption{\ro\ HRI (0.1--2.4 keV) light curve of \xf47\ from July 27 to August 8,
1994. Times of eclipses are indicated using eclipse center HJD~245\,3997.476 and 
orbital period 1.732479~d.
\label{fig:rosat_lc}}
\end{figure*}

\begin{figure*}
\includegraphics[width=10cm,clip]{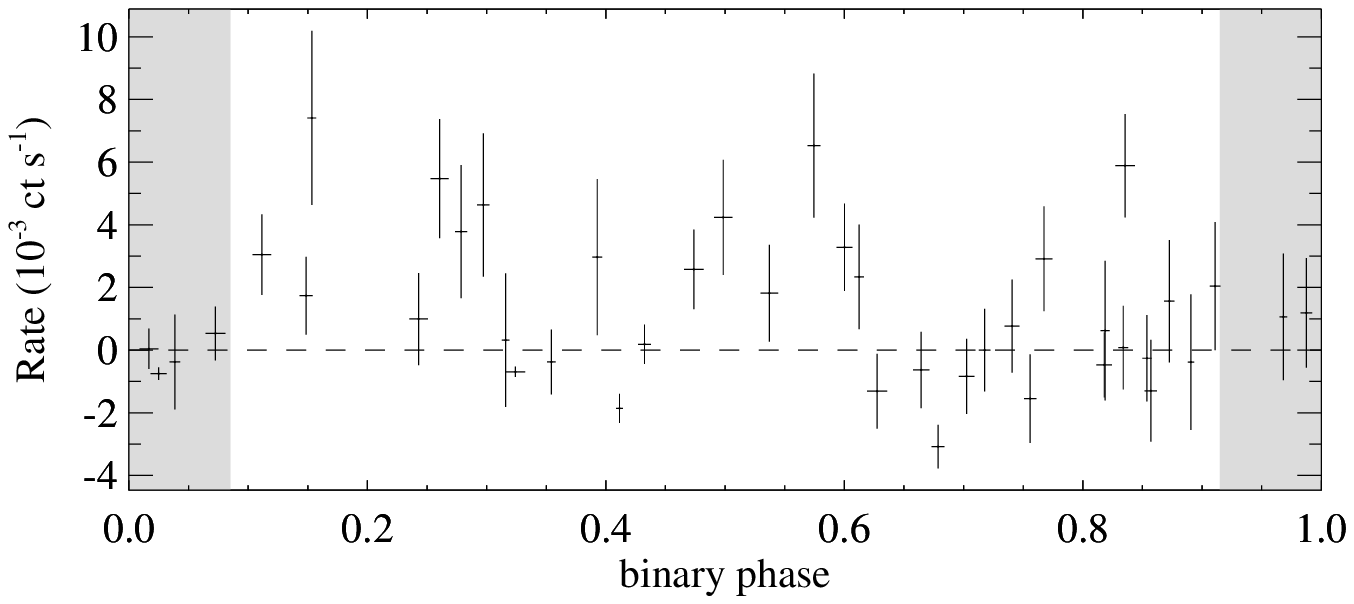}
\caption{\ro\ HRI (0.1--2.4 keV) count rate of \xf47\ versus binary phase
(eclipse center HJD~245\,3997.476 and 
orbital period 1.732479~d, see text). Phase of eclipse is indicated.
\label{fig:rosat_phase}}
\end{figure*}

The \ro\ HRI observations are spread over 12 days and cover several orbital 
periods of \x47 (Fig.~\ref{fig:rosat_lc}). While during the first half of the
observations \x47\ was detected most of the time, the rates are consistent with
zero later on. A \ro\ HRI count rate of $1.0\times 10^{-3}$ \ct\ corresponds to
an un-absorbed luminosity in the 0.2--4.5 keV band of 9\ergs{36} assuming an
absorbed power law spectrum as observed during \chandra\ ObsID 6387 ($N_{\rm H} = 6$\hcm{20}, photon index
$\Gamma = 0.77$). Therefore, during the \ro\ bright state, the \x47\ luminosity was
$\sim5$\ergs{37}, about a factor of two brighter than during the 
\chandra\ and \xmm\ bright state detections. In 
Fig.~\ref{fig:rosat_lc} we mark the extrapolated times of eclipse
using binary ephemeris derived from the \chandra\ and \xmm\ data. 
The \ro\ count rates are plotted versus binary phase in 
Fig.~\ref{fig:rosat_phase}. 
The plot shows intensities compatible with zero at 
phases below 0.1 and above 0.85 indicating that the extrapolation of the
ephemeris by 2560 periods still gives acceptable results. The extrapolation 
of the eclipse egress to the \ro\
observations leads to a phase uncertainty of
0.04. Unfortunately, the
sampling of the phase space by the \ro\ observations during the bright time is
not dense enough to determine an accurate eclipse egress which would allow us to 
improve on the binary ephemeris.

\section{Optical counterpart}
The improved position of \x47\ (see Fig. \ref{fig:fc}) suggests the star
LGGS J013236.92+303228.8 \citep[][]{2006AJ....131.2478M} to be the
optical counterpart listed with $21.011\pm0.011$ mag in $V$ with colors  
$B-V$ of $-0.122\pm0.014$ mag, $U-B$ of $-0.964\pm0.012$, 
$V-R$ of $-0.062\pm0.018$,
and $R-I$ of $0.033\pm0.014$. 

This star was not identified as a variable in the \m33\ CFHT variability survey
\citep[][]{2006MNRAS.371.1405H}. To search for any small optical modulation, we
derived the light curves of this star in the Sloan $g'$, $r'$, and $i'$ bands 
from the CFHT images using the procedure described in \citet[][see
Sect.~\ref{sec:obs}]{2006MNRAS.371.1405H}. Measured magnitudes are given in 
Table~\ref{tbl:opt_lc}.

\begin{figure*}
\includegraphics[width=10cm,clip]{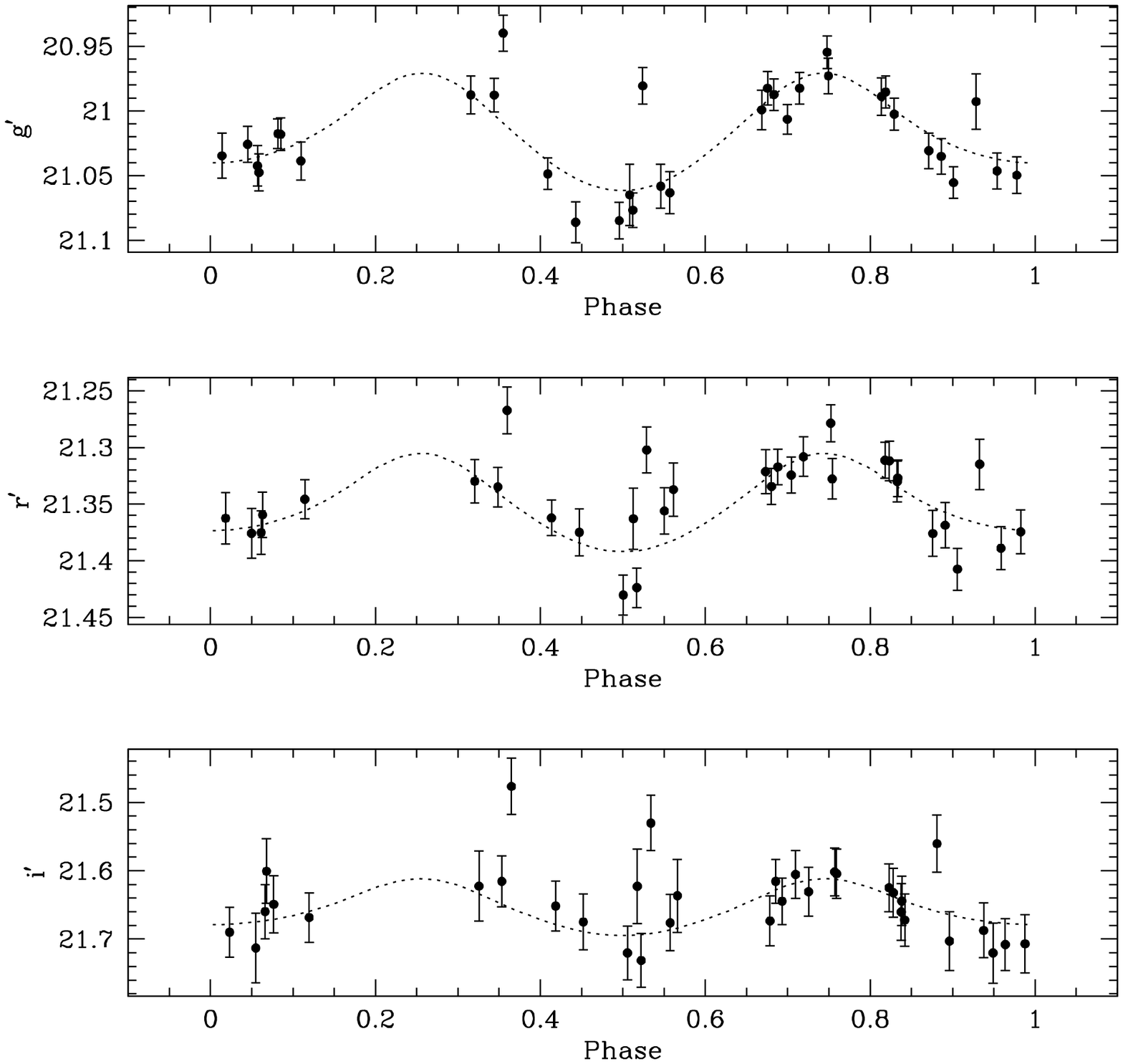}
\caption{Folded light curves, in magnitude, in (top to bottom) $g'$, $r'$ and
$i'$. All light curves are folded using the X-ray ephemeris. The dotted line 
shows a PHOEBE fit for inclination 80$\degr$ and effective temperature of 
33000 K.
\label{fig:opt_lc}}
\end{figure*}

Folding with the X-ray period the 34, 33, and 35 measurements in $g'$, $r'$, 
and $i'$ bands, respectively, reveals
a clear modulation. 
The folded light curves (see Fig.~\ref{fig:opt_lc}) are of
double sinusoidal shape, with one of the minima at the phase of mid X-ray 
eclipse, suggesting an ellipsoidal modulation of a high mass optical companion.
The detection of an optical modulation at the X-ray period confirms the
identification of the optical counterpart.

\begin{figure}
\includegraphics[width=8cm,clip]{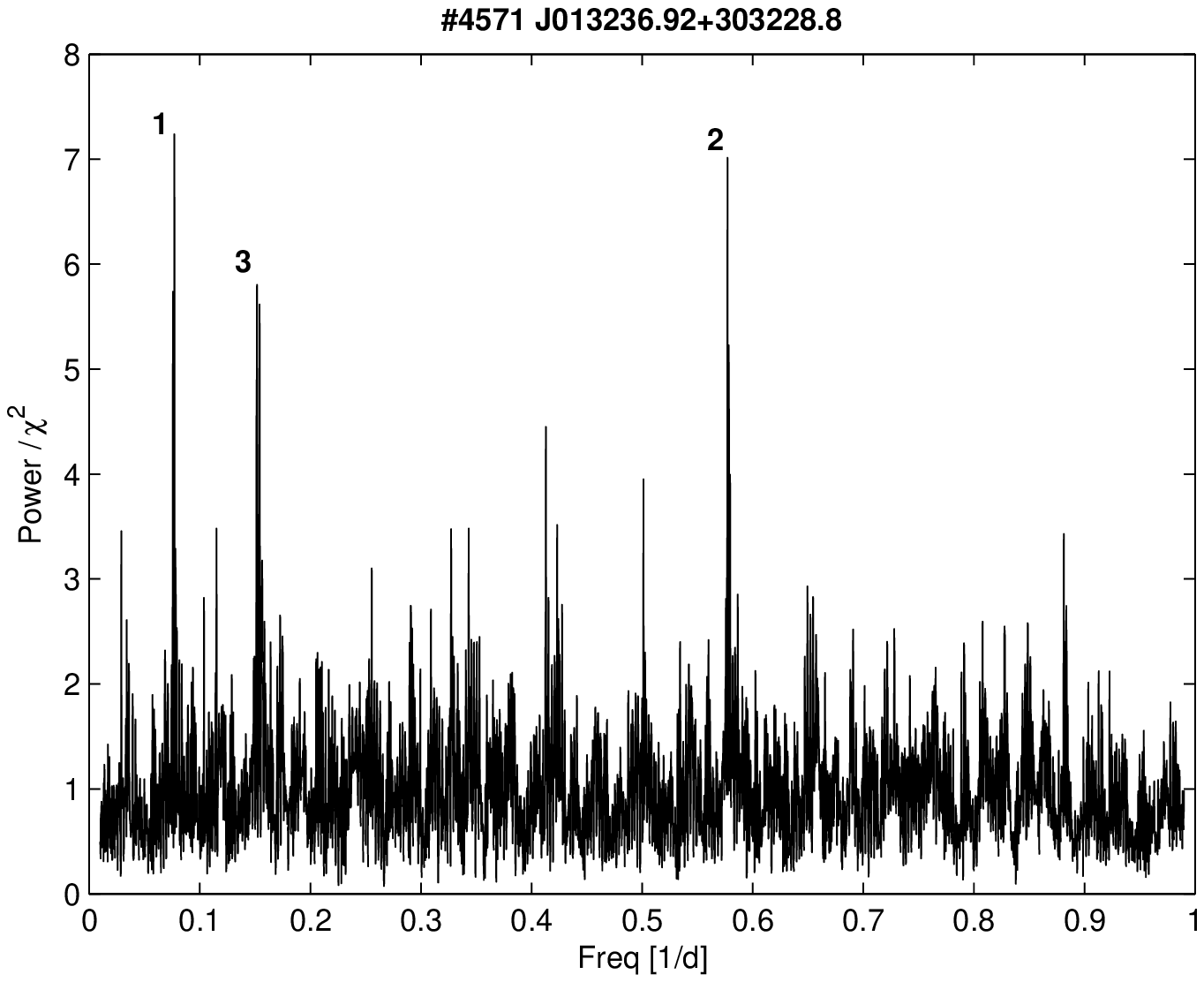}
\caption{Multi-band double-harmonic periodogram after removing outliers 
around phase 0.35 and 0.55. Peak 2 corresponds to the X-ray period (see text).
\label{fig:opt_per}}
\end{figure}

We independently searched the optical data for periodicities. 
We constructed periodograms using the method described at 
\citet[][ Sect. 3.3]{2006MNRAS.370.1429S}
where for each frequency, each band was
fitted with two harmonics and a zero point. The periodogram was taken to be
the amplitudes sum, in quadrature, divided by the $\chi^2$. 
When using all data points the 1.7325 day peak
is the highest, but there are many other similar strong peaks. When
removing the data for the two outliers close to phase 0.35 and phase 0.53
(see Fig.~\ref{fig:opt_lc}), there are only three significant
peaks (0.0771 d$^{-1}$, 0.5772 d$^{-1}$, 0.1517 d$^{-1}$), 
and the 1.7325 day peak is the second strongest (Fig.~\ref{fig:opt_per}).
We checked if there is a justification for
removing these points from the light curve based on sky conditions in the
individual nights or image artifacts.  While we can't guarantee that
there isn't some non-obvious systematic variation, we are inclined to believe
that the variations are real.

\section{Modeling the system}
In determining the parameters of the \x47\ system, we first search for a stellar
model (identified by the effective temperature $T_{\rm eff}$, 
the logarithm of the
surface gravity $g$, and the chemical composition Fe/H in solar units)
that fits the broad-band photometry of the system, assuming the optical
brightness is coming from the optical star alone. 
From the stellar radius, the observed X-ray eclipse width and the periodic
elliptical modulation in the optical, we can derive the two masses as a function
of the orbital inclination. This analysis is similar to the one presented by
\citet[][]{2007A&A...462.1091S} for \xe7. 

\subsection{Optical companion}
To estimate the radius of the optical companion we compared the UBVRI
photometry for the star from \citet[][]{2006AJ....131.2478M} 
to the [Fe/H] = 0 table
of model stellar atmosphere broad-band colors computed by 
\citet[][]{1998A&A...333..231B}. 
Assuming a photometric uncertainty of 0.05 mag, color excess ratios of 
$E(U-B)/E(B-V) = 0.71$, $E(V-R)/E(B-V) = 0.57$ and $E(R-I)/E(B-V) = 0.74$ 
\citep[][]{1998A&A...333..231B}, and a minimum extinction of $E(B-V) = 0.07$ 
which is the foreground reddening to \m33\ from \citet{2000PASP..112..529V}, 
we place a 95\% lower limit on
the effective temperature of 19000 K and a lower limit on $\log(g)$ of
2.5. The formal best fit model has $T_{\rm eff} = 33000$ K, $\log(g) = 4.5$ 
and $E(B-V) = 0.193$ with $\chi^{2} = 2.4$.
Note that the
extinction is in good agreement with the range of values predicted from
analyzing the X-ray spectrum ($0.12 \la E(B-V) \la 0.28$). 
Assuming a distance of 795 kpc to \m33\ and the ``standard value" for the 
ratio of total to selective extinction
$R_v = A_{\rm V}/E(B-V) = 3.1$ \citep[][]{1989ApJ...345..245C}, 
we place a 95\% upper limit of 10.8 \mbox{$\:R_{\sun}$} on the radius.
For the best fit model the radius is 8.0 \mbox{$\:R_{\sun}$}.
For the assumed
distance and best-fit extinction the absolute magnitude of the star is
$M_{V} \sim -4.1$ which, assuming the star is a main-sequence dwarf,
would correspond to a star of type B0V \citep[][]{1982Schmidt}.
We note
that the effective temperature for such a star, $T_{eff} \sim 30000$ K, 
is close to the formal best-fit temperature from modelling the
broad-band colors.

\subsection{Binary system parameters}\label{sec:bin}
\begin{figure}
\includegraphics[width=8cm,clip]{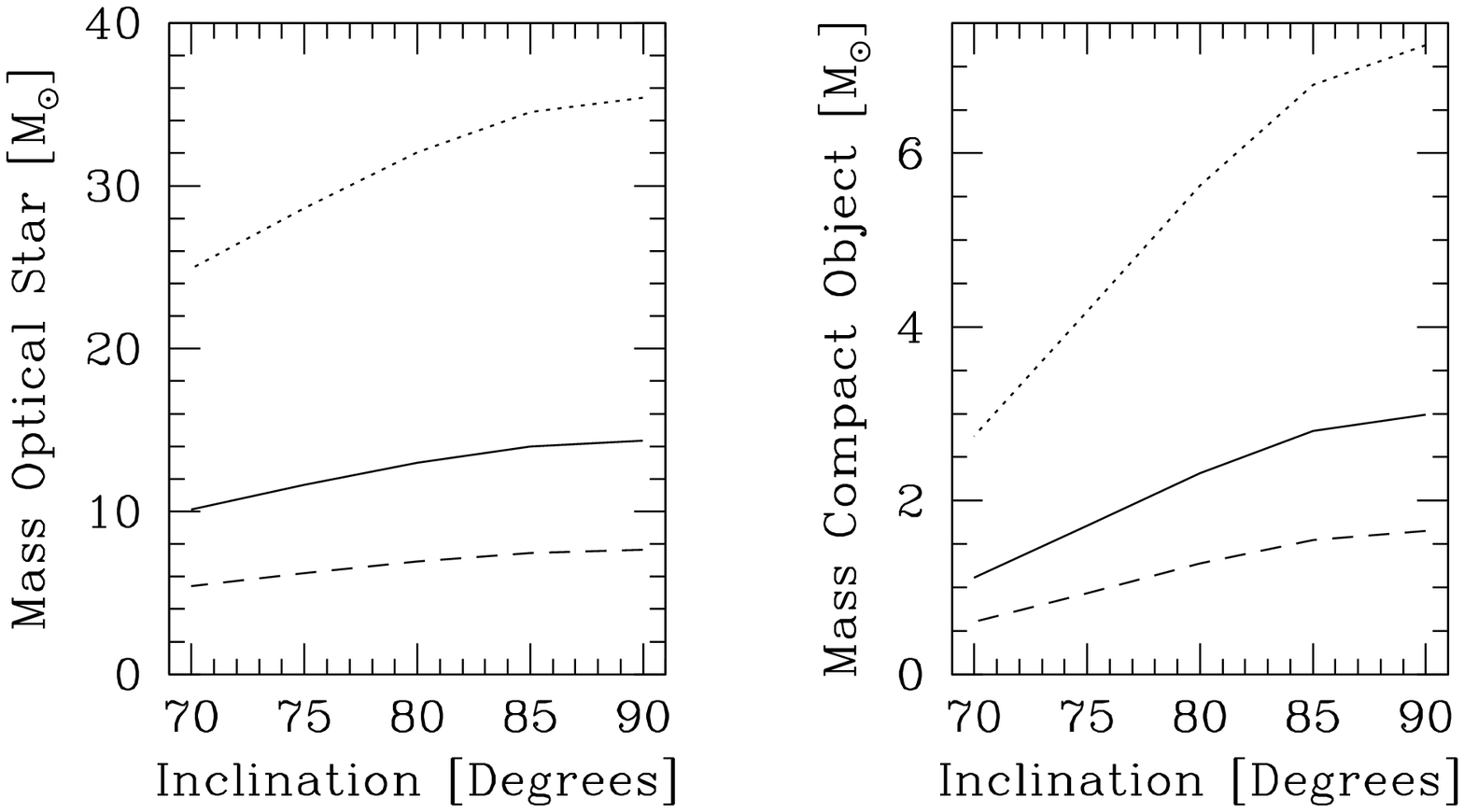}
\caption{Plots showing the mass vs. inclination for three values of
  the stellar temperature. The left plot is for the optical component, the
  right plot is for the compact object.  The solid, the dotted and the dashed
  lines show the results for $T_{\rm eff} = $33000, 19000, and 50000 K, respectively, 
  using the PHOEBE code.
\label{fig:Mvi}}
\end{figure}

We used the "PHysics Of Eclipsing BinariEs (PHOEBE)" program 
of \citet{2005ApJ...628..426P}, a front-end code for the Wilson-Devinney
program \citep{1971ApJ...166..605W,1979ApJ...234.1054W,1990ApJ...356..613W},
to model the obtained
periodic light curves and to derive an estimate of the masses of both
components.  The main effect is due to the
tidally induced ellipsoidal shape of the optical component.  
We ran PHOEBE on the CFHT $g'$, $r'$, and $i'$ light curves, assuming the X-ray
determined eclipse half angle value of $30.6\degr \pm 1.2\degr$ and using 
linear limb darkening coefficients from \citet{2004A&A...428.1001C},
stepping through inclination and fitting for the mass ratio.
Figure~\ref{fig:opt_lc} shows a representative PHOEBE model fit 
to the light curves
for a binary system with an inclination of $i = 80 \degr$. We note that 
due to
the degeneracy between $q$, $i$ and the Roche-lobe-filling factor, the
model fit is similar for other allowed inclinations.
Figure~\ref{fig:Mvi} 
shows a plot of the optical star and compact-object masses
as a function of $i$ for each of these different $T_{\rm eff}$ values.
Below $i = 70\degr$ the optical star fills its Roche Lobe.
From the plot for the compact object mass, 
a neutron star with canonical mass of 
1.4 \mbox{$\:M_{\sun}$} seems to be allowed for inclinations below 85\degr\ for the 
50000 K model, but only for below 75\degr\ for the 33000 K model. For the 
19000 K model, a canonical mass neutron star seems to be excluded. 
A neutron star with canonical mass at the best fit companion temperature 
$T_{\rm eff}$ of 33000 K would
imply a system inclination $i = 72 \degr$ and a companion mass of 
10.9 \mbox{$\:M_{\sun}$}. 

\section{Discussion and summary}
The time variability of \x47\  in the \cm33\ observations can be explained
with a 1.7 d eclipsing X-ray binary with an eclipse half angle of $30.6\degr$.
The identification with an OB star in \m33\ that shows an ellipsoidal light
curve, clearly establishes the system as the second HMXB in \m33. 
We do not detect
pulsations in the X-ray flux and therefore the compact object in the system  could be
a neutron star or a black hole. The hard X-ray spectrum, however, may indicate
a neutron star system \citep[see e.g.][]{1983ApJ...270..711W}. 

We can not exclude that the long term variability of the system in the \chandra, 
\xmm\ and \ro\
observations can be of irregular nature. However, it also could reflect a long
term periodicity similar to those known in Her X--1, LMC X--4 or SMC X--1. 
The \chandra\ 
bright state observations (ObsID 6387 and 6385) are separated by 84 days. With
our sparse unsystematic monitoring, we can not decide if the source had another
on-state in between as would be expected by extrapolating from the 35 d period 
of Her X--1, a system with a similar orbital period.

Inspecting the available \citet[][]{2006AJ....131.2478M} colour and 
emission-line images we find that the object is located in a region with few 
(bright) stars. Also, there are no close-by emission nebulae, indicating that 
the object is older than \mx7\ which is still located in a \HII\ region.

\x47\ was serendipitously detected far off axis in the \cm33\ observations.
Only with the help of dedicated deep observations with good time resolution 
it will be possible
to characterize long term variability of \x47\ and to search for the expected
pulsation period below 50 seconds.  

\acknowledgments
Support for this work was provided by the National Aeronautics 
and Space Administration through Chandra Award Number G06-7073A 
issued by the Chandra X-ray Observatory Center, which is operated 
by the Smithsonian Astrophysical Observatory for and on behalf of 
the National Aeronautics Space Administration under contract 
NAS8-03060. PPP, and TJG acknowledge support under 
NASA contract NAS8-03060.
This work is partly based on observations obtained with \xmm, an ESA science mission 
with instruments and contributions directly funded by ESA Member States and
NASA. The \xmm\ project is supported by the Bundesministerium f\"{u}r
Wirtschaft und Technologie / Deutsches Zentrum f\"{u}r Luft- und Raumfahrt 
(BMWI/DLR FKZ 50 OX 0001) and the Max-Planck Society.
This research has made use of data obtained through the High Energy Astrophysics
Science Archive Research Center Online Service, provided by the NASA/Goddard 
Space Flight Center.
We used data
of observations obtained with the MegaPrime/MegaCam, a joint project of CFHT and
CEA/DAPNIA, at the Canada-France-Hawaii Telescope (CFHT) which is operated by
the National Reasearch Council (NRC) of Canada, the Institute National des
Sciences de l'Univers of the Centre National de la Recherche Scientifique of
France, and the University of Hawaii.
We used images of the USNOFS Image and Catalogue Archive
operated by the United States Naval Observatory, Flagstaff Station
\footnote{http://www.nofs.navy.mil/data/fchpix/}.
This work has made use of {\tt ACIS-Extract}\footnote{\tt
http://www.astro.psu.edu/xray/docs/TARA/ae\_users\_guide.html},  SAOImage 
DS9\footnote{\tt http://hea-www.harvard.edu/RD/}, developed 
by the Smithsonian Astrophysical Observatory \citep{2003ASPC..295..489J}, 
the {\tt XSPEC}\footnote{\tt http://xspec.gsfc.nasa.gov/}
spectral fitting package \citep{1996ASPC..101...17A}, the 
FUNTOOLS\footnote{\tt http://hea-www.harvard.edu/RD/funtools} 
utilities package, the HEASARC 
FTOOLS\footnote{\tt 
http://heasarc.gsfc.nasa.gov/docs/software/lheasoft/ftools/}
package, and the 
CIAO\footnote{\tt http://cxc.harvard.edu/ciao/}
(\chandra{} Interactive Analysis of Observations) package.
The anonymous referee is acknowledged for her/his constructive comments and
suggestions.

{\it Facilities:} \facility{CXO (ACIS)}, \facility{XMM (EPIC)},
\facility{ROSAT (HRI)}.

\clearpage
\begin{table*}
\footnotesize
\caption{\chandra\ observations of the \cm33\ program covering [PMH2004]~47.
\label{tbl:obs_chandra}}
\begin{tabular}{rrrrrrrrr}
\tableline\tableline
\cm33 & Obs.id. & Obs.dates\tablenotemark{a} & On-time & ACIS & Offax &  
\multicolumn{2}{c}{\xf47\ binary} \\
Field & &  & (ks) & CCD-ID & (') & \multicolumn{1}{l}{phase\tablenotemark{b}} &
cycle\tablenotemark{b} &comment\\
\tableline
  & 1730 & 2000-07-12 &  40\tablenotemark{c} & 7 & 17.6 & 0.795-1.062 & -1305,-1304&ingress, off \\
4 & 6382 & 2005-11-23 &  73 & 0 & 10.2 & 0.168-0.589 &  -173      &faint  \\
4 & 7226 & 2005-11-26 &  25 & 0 & 10.2 & 0.703-0.856 &  -172      &faint  \\
4 & 6383 & 2006-06-15 &  99 & 3 & 10.8 & 0.754-1.424 &  -56,-55   &eclipse, bright  \\
6 & 6387 & 2006-06-26 &  78 & 6 & 19.7 & 0.993-1.507 &  -50,-49   &egress, bright  \\
6 & 7344 & 2006-07-01 &  22 & 6 & 19.7 & 0.882-0.998 &  -47       &ingress, off  \\
5 & 6385 & 2006-09-18 &  91 & 2 & 10.9 & 0.534-1.139 &  -1,0      &eclipse, bright  \\
1 & 6377 & 2006-09-25 &  94 & 6 & 17.4 & 0.906-1.507 &  3,4       &egress, faint \\
\tableline
\end{tabular}
\tablenotetext{a}{date of start of observation}
\tablenotetext{b}{with respect to eclipse center HJD~245\,3997.476 and 
orbital period 1.732479~d
(see text)}
\tablenotetext{c}{residual exposure after screening first 10 ks due to
background flaring}
\end{table*}

\begin{table*}
\footnotesize
\caption{\xmm\ observations covering [PMH2004]~47.
\label{tbl:obs_xmm}}
\begin{tabular}{rrrrrrrr}
\tableline\tableline
Obs.id. & Obs.dates\tablenotemark{a} & On-time & Offax 
& $L_{\rm X}$ \tablenotemark{b}&  \multicolumn{2}{c}{\xf47\ binary} \\
 &  & (ks)  & (') & &\multicolumn{1}{l}{phase\tablenotemark{c}} &
cycle\tablenotemark{c} &comment\\
\tableline
0102640401 & 2000-08-02 &  16  &  5.6 & $<33$ & 0.952-1.039 & -1293,-1292&eclipse \\
0102640501 & 2001-07-05 &  12  & 10.7 & $836\pm65$ & 0.729-0.809 &  -1098    &edge of pn,bright  \\
0102641001 & 2001-07-08 &  12  & 14.7 & $<206$ & 0.992-1.072 &  -1097,-1096    &eclipse  \\
0102641101 & 2001-07-08 &  12  &  3.2 & $1732\pm68$ & 0.099-0.179 &  -1096   &bright \\
0102642301 & 2002-01-27 &  13  & 13.7 & $<67$ & 0.305-0.385 &  -979   &off  \\
0141980601 & 2003-01-23 &  14  &  5.8 & $<13$ & 0.919-1.011 &  -771,-770 &eclipse  \\
0141980101 & 2003-07-11 &  16  & 14.7 & $<205$ & 0.377-0.484 &  -673    &off  \\
\tableline
\end{tabular}
\tablenotetext{a}{date of start of observation}
\tablenotetext{b}{0.2--4.5 keV absorption corrected luminosity or 3$\sigma$ 
upper limits in units of \oergs{34}
assuming an absorbed power law spectrum ($N_{\rm H} = 6$\hcm{20}, photon index
$\Gamma = 1.7$) and a distance to \m33\ of 795 kpc 
\citep[][]{1991PASP..103..609V} which we use
throughout the paper}
\tablenotetext{c}{with respect to eclipse center HJD~245\,3997.476 and 
orbital period 1.732479~d
(see text)}
\end{table*}

\begin{table*}
\caption{Power law spectral modeling results for \xf47\ for the bright time in the 
\chandra\ ACIS-I ObsIDs 6385 and 6387
and for the \xmm\
EPIC observation 0102641101 (all EPIC instruments fitted together). 
For each instrument, we give the effective integration time $t_{\rm int}$ and
the raw count rate. The number of energy bins reduced by the number of free
parameters defines the degrees of freedom $\nu$.
90$\%$ errors are given. 
\label{tbl:spectra}}
\begin{tabular}{llrrrrrrr}
\tableline\tableline
Observation & Inst. & $t_{\rm int}$ & Rate\tablenotemark{a}  
     & $N_{\rm H\,M33}$ \tablenotemark{b}& 
     $\Gamma$& $L_{\rm X}$\tablenotemark{c} & $\nu$& $\chi^2/\nu$\\
&& (ks) & & &  & &  &  \\
\tableline
\chandra\ 6385&ACIS I& 63.1 & 2.27 & $16_{-12}^{+15}$ &  
$0.94\pm0.13$&  $1.9$  & 66 &1.38\\
\noalign{\smallskip}
\noalign{\smallskip}
\noalign{\smallskip}
\chandra\ 6387&ACIS I& 62.5 & 2.20 & $<8$ &  
$0.77_{-0.10}^{+0.12}$&  $2.0$  & 77 &1.19\\
\noalign{\smallskip}
\noalign{\smallskip}
\noalign{\smallskip}
\xmm &PN& 8.8 &6.40 & $1.2_{-1.2}^{+8.5}$ &   
$0.88\pm0.13$&  $1.8$  & 28   & 1.38\\ 
EPIC &MOS1& 9.8&1.73 &  &&       
&&\\
0102542301&MOS2& 9.8&1.77 &  & $$&$$ 
& $$  &  \\ 
\tableline
\end{tabular} 
\tablenotetext{a}{Raw count rate in units of $10^{-2}$ ct s$^{-1}$ 
as given in XSPEC}
\tablenotetext{b}{Absorption in units of $10^{20}$~cm$^{-2}$ 
exceeding the fixed Galactic foreground of
6.0\hcm{20}}
\tablenotetext{c}{X-ray luminosity in the 0.2--4.5 keV band in units of 
$10^{37}$ erg s$^{-1}$, corrected for absorption,
for extraction radii, and for vignetting}
\end{table*}   

\begin{table*}
\caption{Optical light curve of the star LGGS J013236.92+303228.8 identified 
with \xf47. Magnitude with error in the $g'$, $r'$, and $i'$ filter and heliocentric 
Julian date (HJD) of the observation are given.
\label{tbl:opt_lc}}
\begin{tabular}{rc|rc|rc}
\tableline\tableline
\multicolumn{1}{l}{HJD $-$} & 
$g'$ &
\multicolumn{1}{l}{HJD $-$} & 
$r'$ &
\multicolumn{1}{l}{HJD $-$} & 
\multicolumn{1}{c}{$i'$} \\
245\,0000.0&(mag)&245\,0000.0&(mag)&245\,0000.0&(mag)\\
\tableline
2873.984375 &  21.077(13) & 2873.992665 &  21.424(17) & 2874.001810 &  21.731(39) \\
2876.041448 &  21.006(12) & 2876.049696 &  21.324(16) & 2876.058663 &  21.606(35) \\
2876.067077 &  20.982(12) & 2876.075340 &  21.308(17) & 2876.086306 &  21.630(36) \\
2881.949358 &  21.039(15) & 2881.957628 &  21.346(17) & 2881.966625 &  21.668(36) \\
2882.917645 &  20.999(15) & 2882.925921 &  21.321(19) & 2882.934887 &  21.673(36) \\
2882.943189 &  20.987(12) & 2882.951458 &  21.317(16) & 2882.960463 &  21.645(34) \\
2886.083231 &  21.085(14) & 2886.091515 &  21.430(18) & 2886.100505 &  21.720(40) \\
2903.847418 &  20.973(14) & 2903.855644 &  21.328(18) & 2903.864595 &  21.604(36) \\
2905.975845 &  21.050(14) & 2905.984073 &  21.374(19) & 2905.993022 &  21.707(43) \\
2908.917034 &  20.982(13) & 2908.925269 &  21.334(16) & 2908.934226 &  21.616(32) \\
2910.026011 &  20.988(15) & 2910.034266 &  21.330(19) & 2910.043261 &  21.622(51) \\
2931.777952 &  21.031(14) & 2931.786238 &  21.376(20) & 2931.795225 &  21.560(42) \\
2931.804198 &  21.035(14) & 2931.812455 &  21.369(20) & 2931.821460 &  21.703(43) \\
2931.829883 &  21.055(12) & 2931.838149 &  21.407(18) & 2931.912971 &  21.720(44) \\
2931.921486 &  21.046(14) & 2931.929751 &  21.389(19) & 2931.938758 &  21.708(38) \\
2940.741715 &  21.026(14) & 2940.749948 &  21.376(22) & 2940.758913 &  21.713(51) \\
2963.800785 &  20.940(14) & 2963.809086 &  21.267(21) & 2963.818136 &  21.477(41) \\
2991.813734 &  20.981(14) & 2991.821989 &  21.302(20) & 2991.830949 &  21.530(40) \\
3240.978670 &  20.988(13) & 3240.986361 &  21.335(17) & 3240.994707 &  21.616(37) \\
3256.073787 &  21.042(16) & 3256.081416 &  21.375(19) & 3256.089708 &  21.660(40) \\
3260.849553 &  20.989(14) & 3260.857172 &  21.311(16) & 3260.865472 &  21.625(35) \\
3261.939754 &  21.086(16) & 3261.947396 &  21.375(21) & 3261.955702 &  21.675(41) \\
3286.845648 &  20.985(12) & 3286.853273 &  21.312(17) & 3286.861569 &  21.632(36) \\
3287.868151 &  21.049(12) & 3286.869976 &  21.330(18) & 3286.878287 &  21.660(42) \\
3296.786986 &  21.063(16) & 3287.876147 &  21.362(16) & 3286.885816 &  21.672(38) \\
3313.022723 &  20.993(21) & 3296.794665 &  21.337(23) & 3287.884760 &  21.652(36) \\
3316.713424 &  21.048(14) & 3313.030339 &  21.315(22) & 3296.803012 &  21.637(53) \\
3316.753196 &  21.018(11) & 3316.721061 &  21.359(20) & 3313.038618 &  21.687(40) \\
3316.759412 &  21.018(13) & 3317.914851 &  21.279(16) & 3316.729355 &  21.601(47) \\
3317.907228 &  20.955(13) & 3322.762627 &  21.356(21) & 3316.744412 &  21.649(42) \\
3322.755010 &  21.058(17) & 3328.770804 &  21.362(23) & 3317.923132 &  21.601(35) \\
3328.763167 &  21.035(17) & 3345.775498 &  21.327(16) & 3322.774943 &  21.676(41) \\
3345.767876 &  21.003(12) & 3386.799264 &  21.363(27) & 3328.779078 &  21.690(37) \\
3386.791633 &  21.065(24) &             &             & 3345.783809 &  21.644(36) \\
            &             &             &             & 3386.807594 &  21.623(54) \\
\tableline
\end{tabular} 
\end{table*}

\clearpage

\begin{figure*}
\centering
\begin{minipage}[t]{10cm}
\vspace{0pt}
\includegraphics[origin=br,height=10cm,bb= 240 25 425 710,angle=-90,clip]{f1a.eps} 
\end{minipage}
\begin{minipage}[t]{2.9cm}
\vspace{0pt}
\hspace{0.1cm} \includegraphics[width=2.9cm,clip]{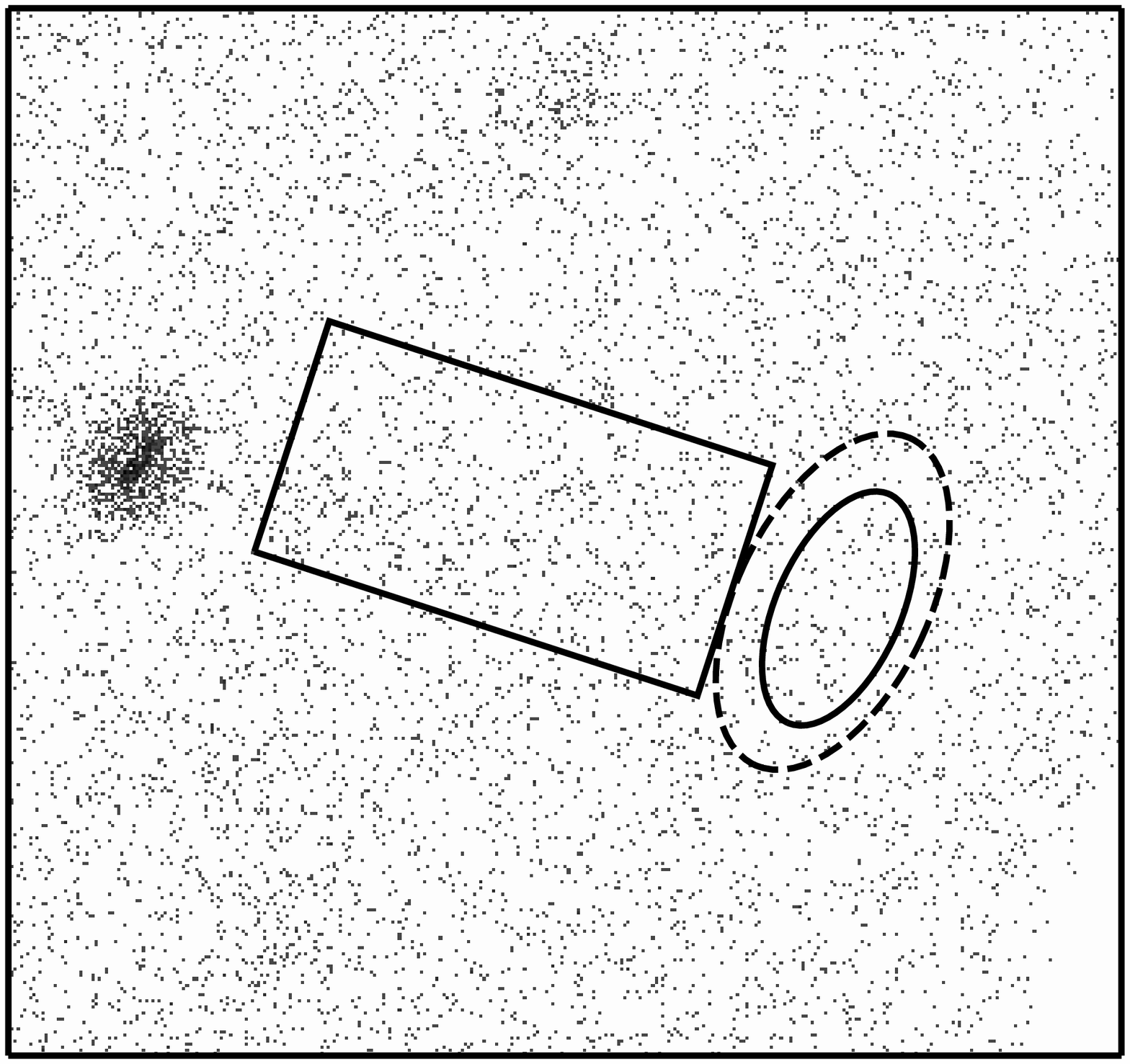}
\end{minipage}
\vfill
\begin{minipage}[t]{10cm}
\vspace{0pt}
\includegraphics[origin=br,height=10cm,bb= 240 25 425 710,angle=-90,clip]{f1c.eps} 
\end{minipage}
\begin{minipage}[t]{2.9cm}
\vspace{0pt}
\includegraphics[width=2.9cm,clip]{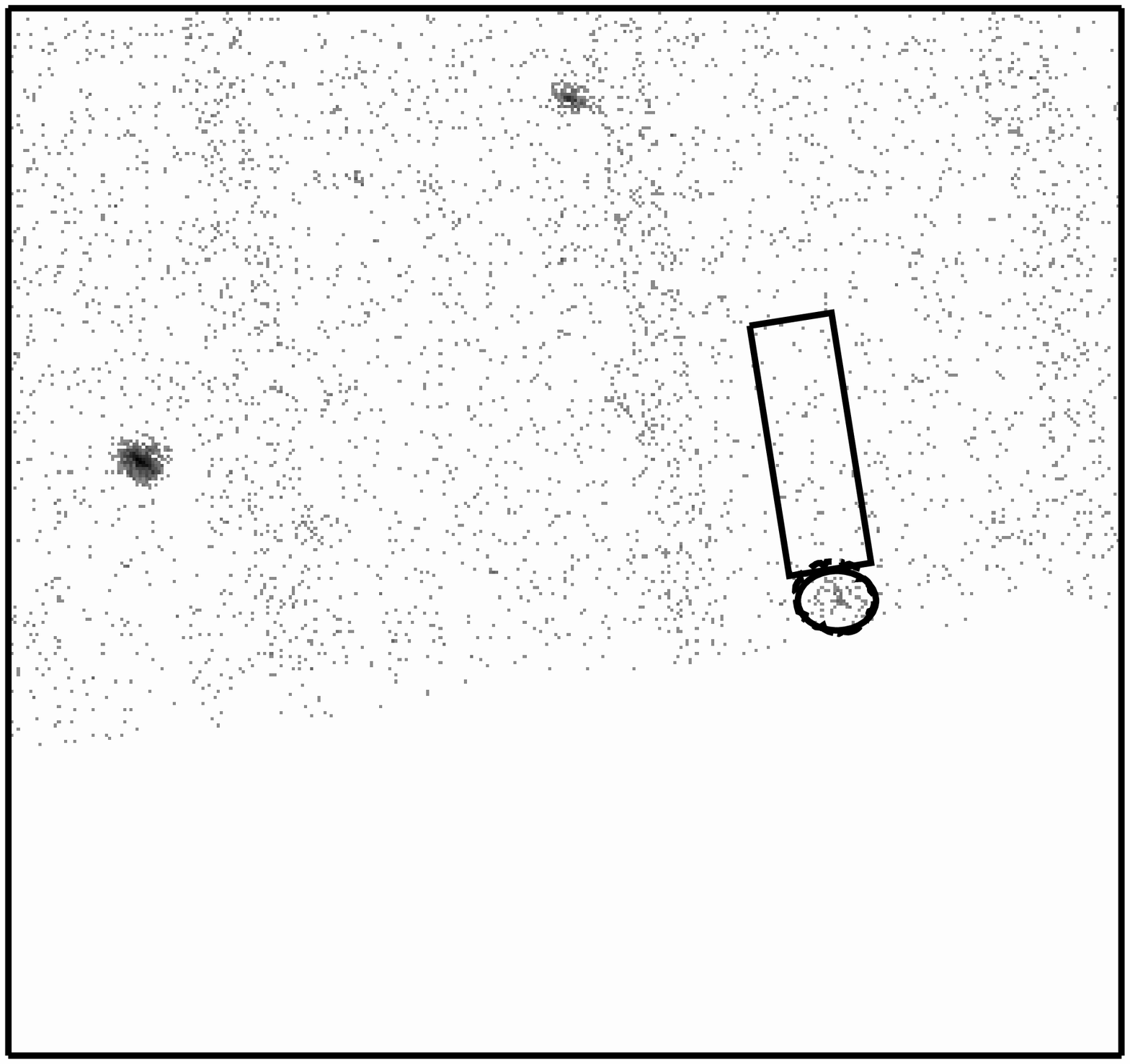}
\end{minipage}
\vfill
\begin{minipage}[t]{10cm}
\vspace{0pt}
\includegraphics[origin=br,height=10cm,bb= 240 25 425 710,angle=-90,clip]{f1e.eps} 
\end{minipage}
\begin{minipage}[t]{2.9cm}
\vspace{0pt}
\includegraphics[width=2.9cm,clip]{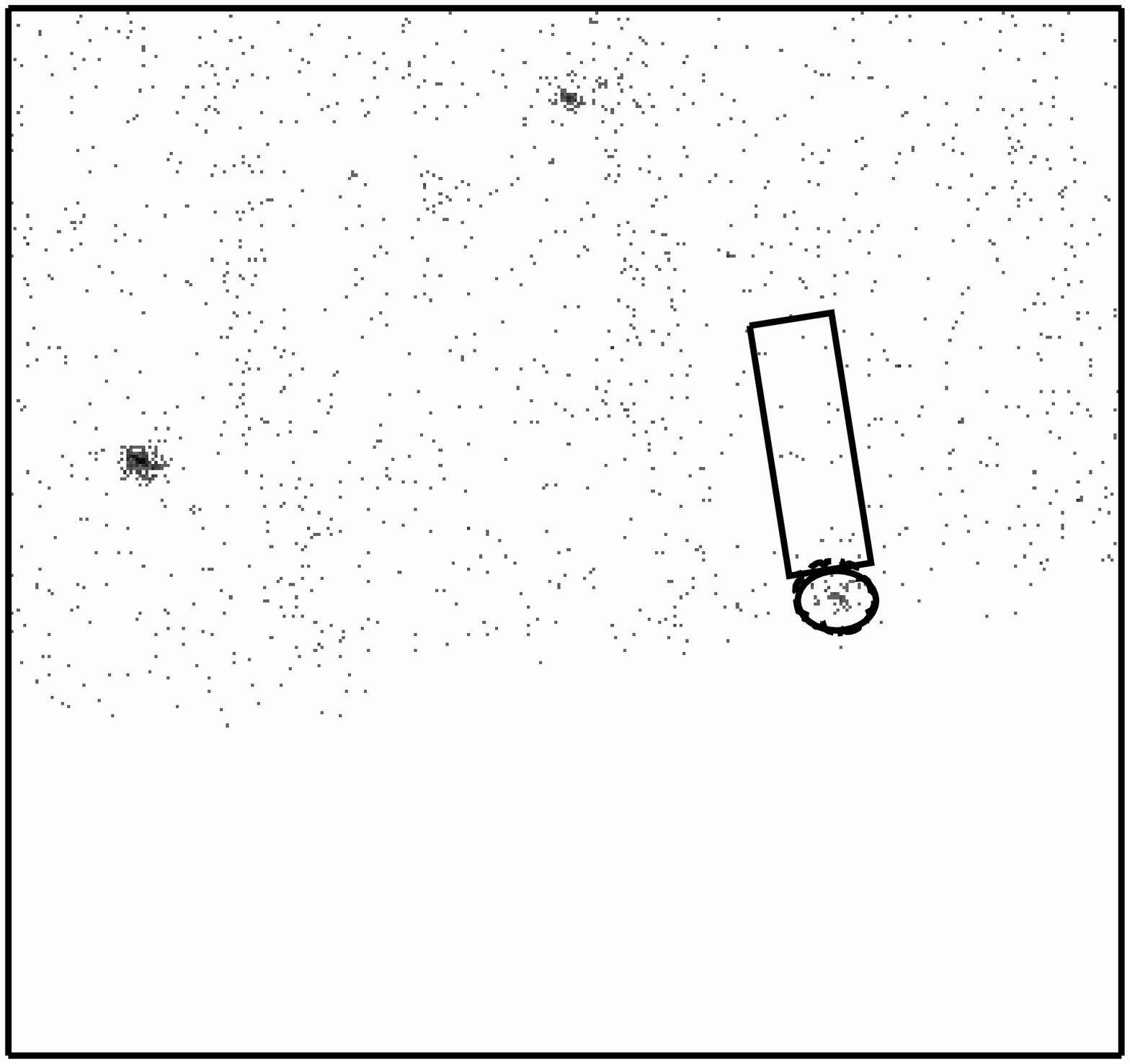} 
\end{minipage}
\vfill
\begin{minipage}[t]{10cm}
\vspace{0pt}
\includegraphics[origin=br,height=10cm,bb= 240 25 475 710,angle=-90,clip]{f1g.eps} 
\end{minipage}
\begin{minipage}[t]{2.9cm}
\vspace{0pt}
\includegraphics[width=2.9cm,clip]{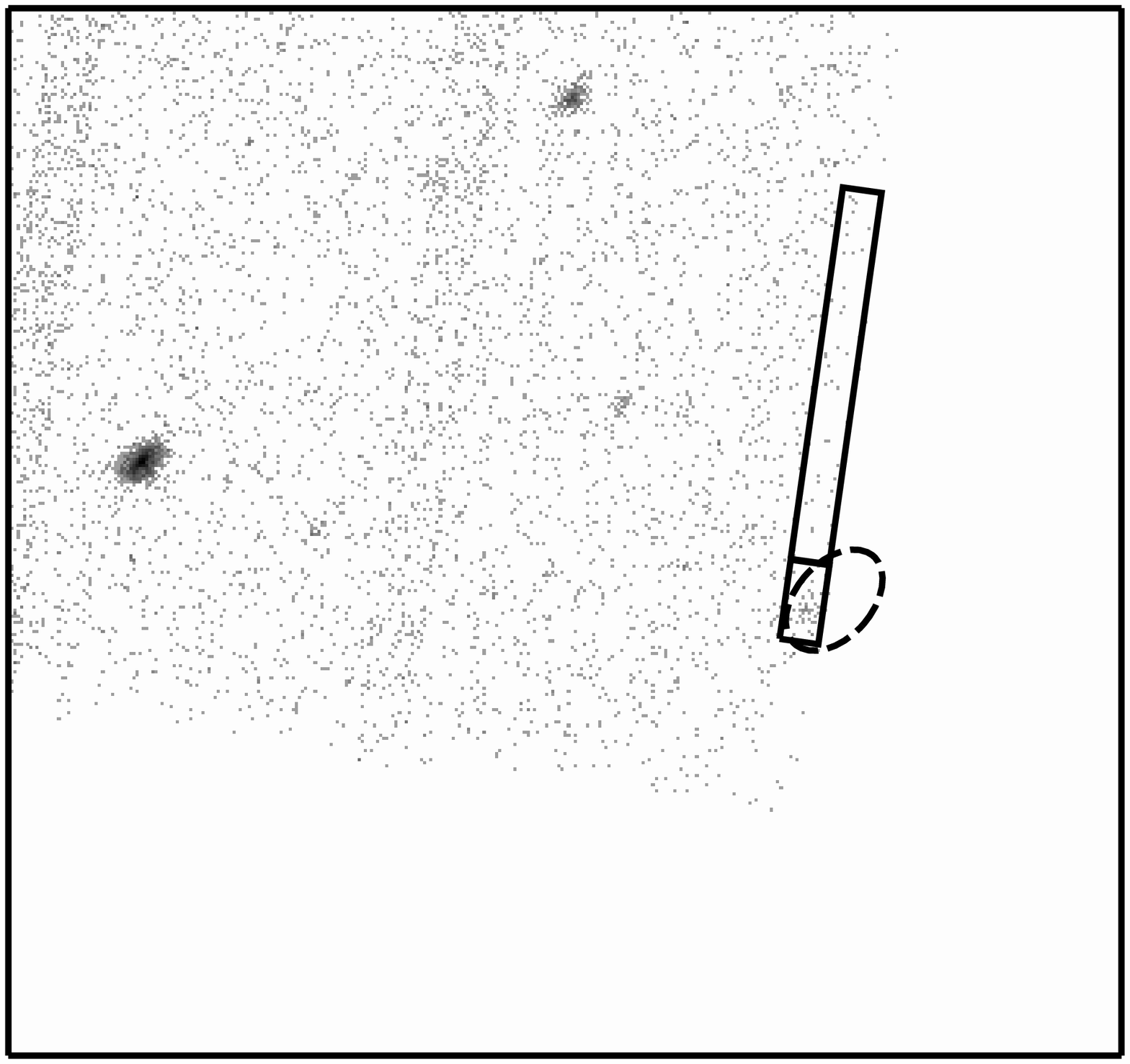}
\end{minipage}
\vfill
\caption{{\bf left:} \chandra\ ACIS-I light curves (0.5--5 keV) of ObsID 1730,
6382, 7226, 6383, 6387, 7344, 6385 and 6377 
integrated over 5000~s, 5000~s, 5000~s, 5000~s, 1000~s, 5000~s, 1000~s and 
3000~s, respectively.
Time zero corresponds to HJD~245\,0000.0 + (1737.8527, 3698.0480, 3700.7070,
3901.7629, 3912.5724, 3917.5767, 3996.6692q and 4004.2427), respectively 
(solar system barycentre corrected). {\bf right:} $6\arcmin\times6\arcmin$ 
ACIS images (0.5--5 keV) of the \xf47\ area for the individual ObsIDs. 
Source and background extraction regions
are indicated by  rectangles and ellipses.  The dashed ellipses
show the HRMA PSF ~90\% enclosed count fraction regions.
\label{fig:lc_chandra}}
\end{figure*}
\clearpage

\begin{figure*}
\centering
\begin{minipage}[t]{10cm}
\vspace{0pt}
\includegraphics[origin=br,height=10cm,bb= 240 25 425 710,angle=-90,clip]{f1i.eps} 
\end{minipage}
\begin{minipage}[t]{2.9cm}
\vspace{0pt}
\includegraphics[width=2.9cm,clip]{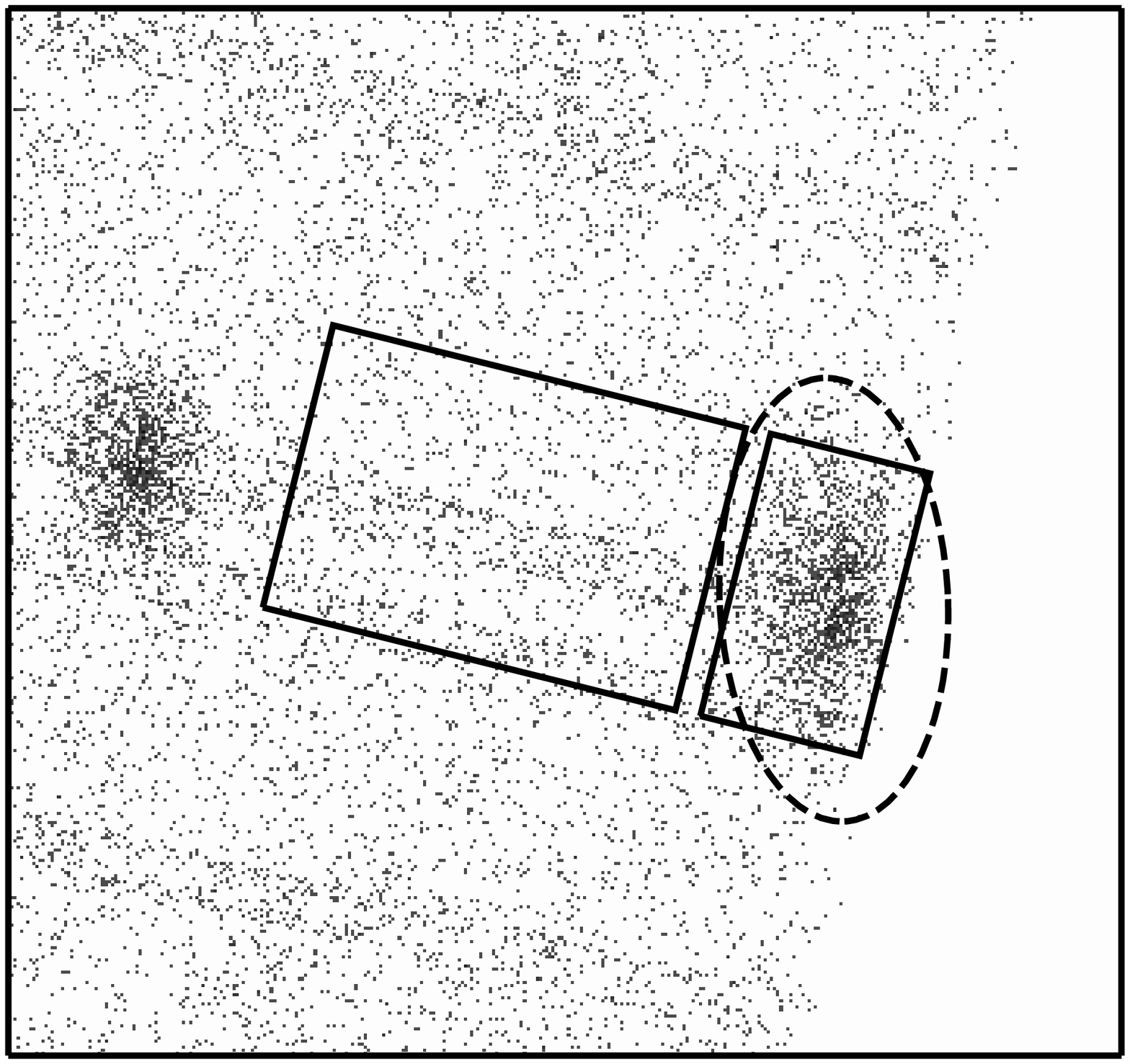} 
\end{minipage}
\begin{minipage}[t]{10cm}
\vspace{0pt}
\includegraphics[origin=br,height=10cm,bb= 240 25 425 710,angle=-90,clip]{f1k.eps} 
\end{minipage}
\begin{minipage}[t]{2.9cm}
\vspace{0pt}
\includegraphics[width=2.9cm,clip]{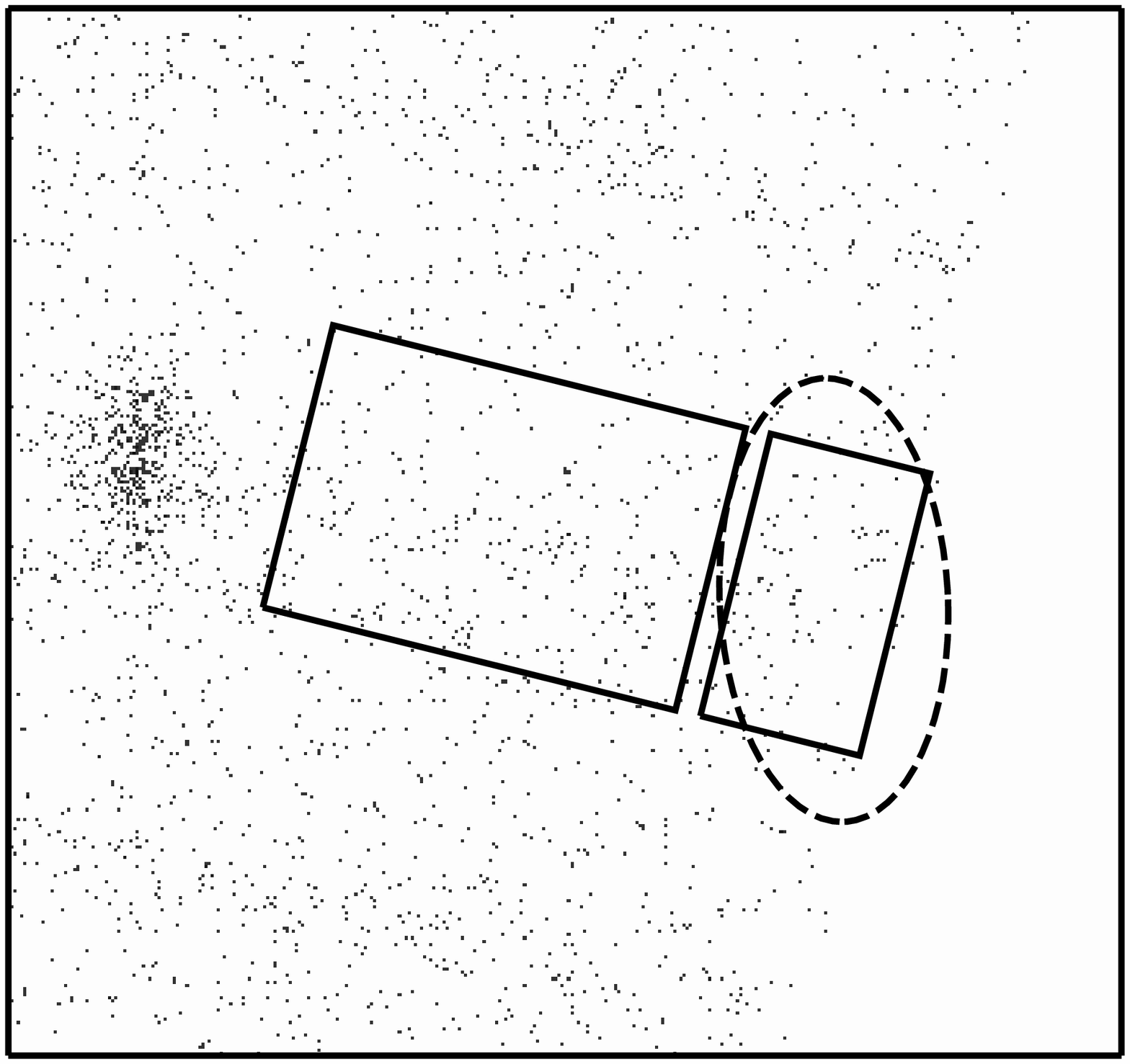}
\end{minipage}
\begin{minipage}[t]{10cm}
\vspace{0pt}
\includegraphics[origin=br,height=10cm,bb= 240 25 425 710,angle=-90,clip]{f1m.eps} 
\end{minipage}
\begin{minipage}[t]{2.9cm}
\vspace{0pt}
\includegraphics[width=2.9cm,clip]{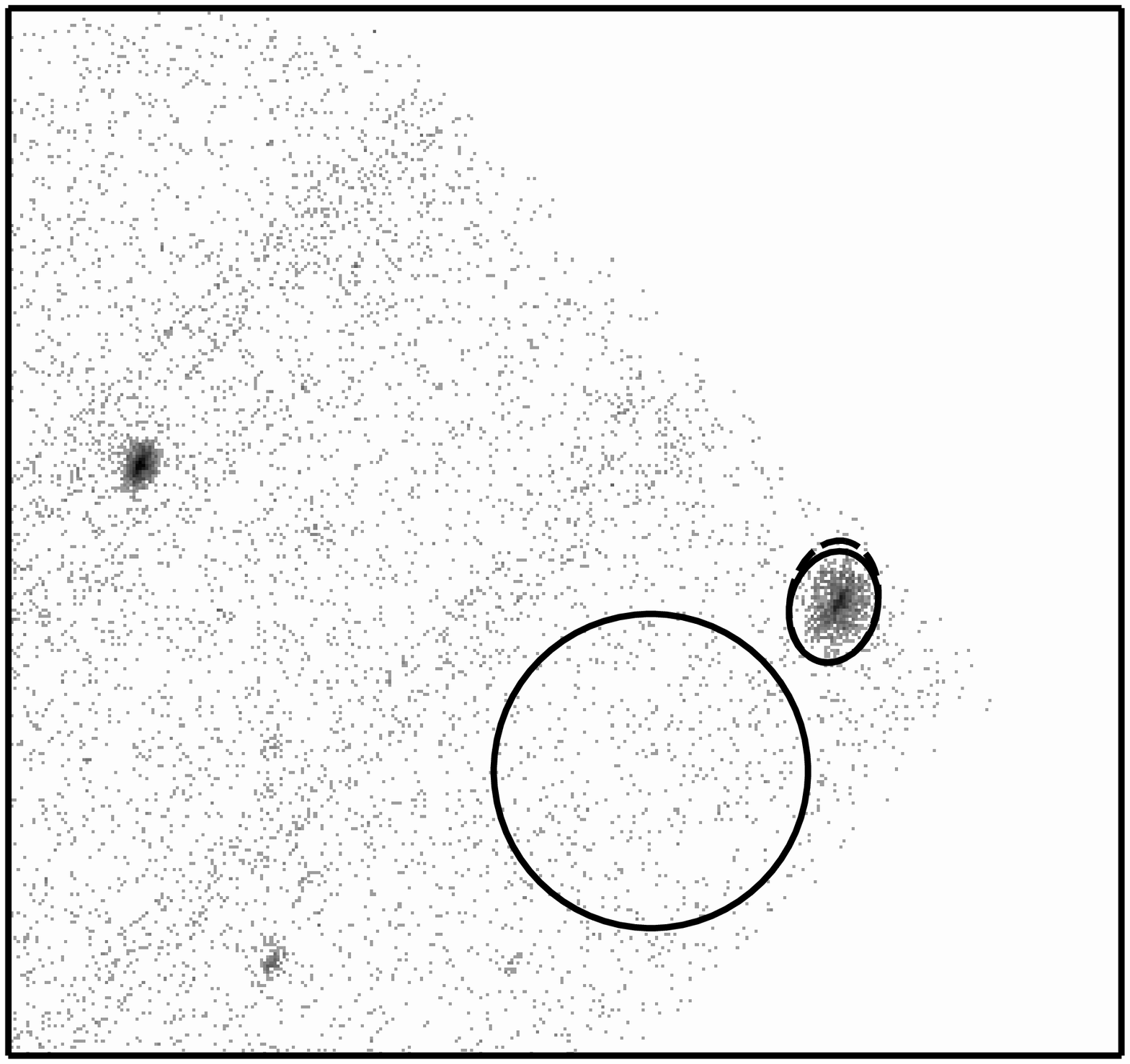}
\end{minipage}
\begin{minipage}[t]{10cm}
\vspace{0pt}
\includegraphics[origin=br,height=10cm,bb= 240 25 475 710,angle=-90,clip]{f1o.eps}
\end{minipage}
\begin{minipage}[t]{2.9cm}
\vspace{0pt}
\includegraphics[width=2.9cm,clip]{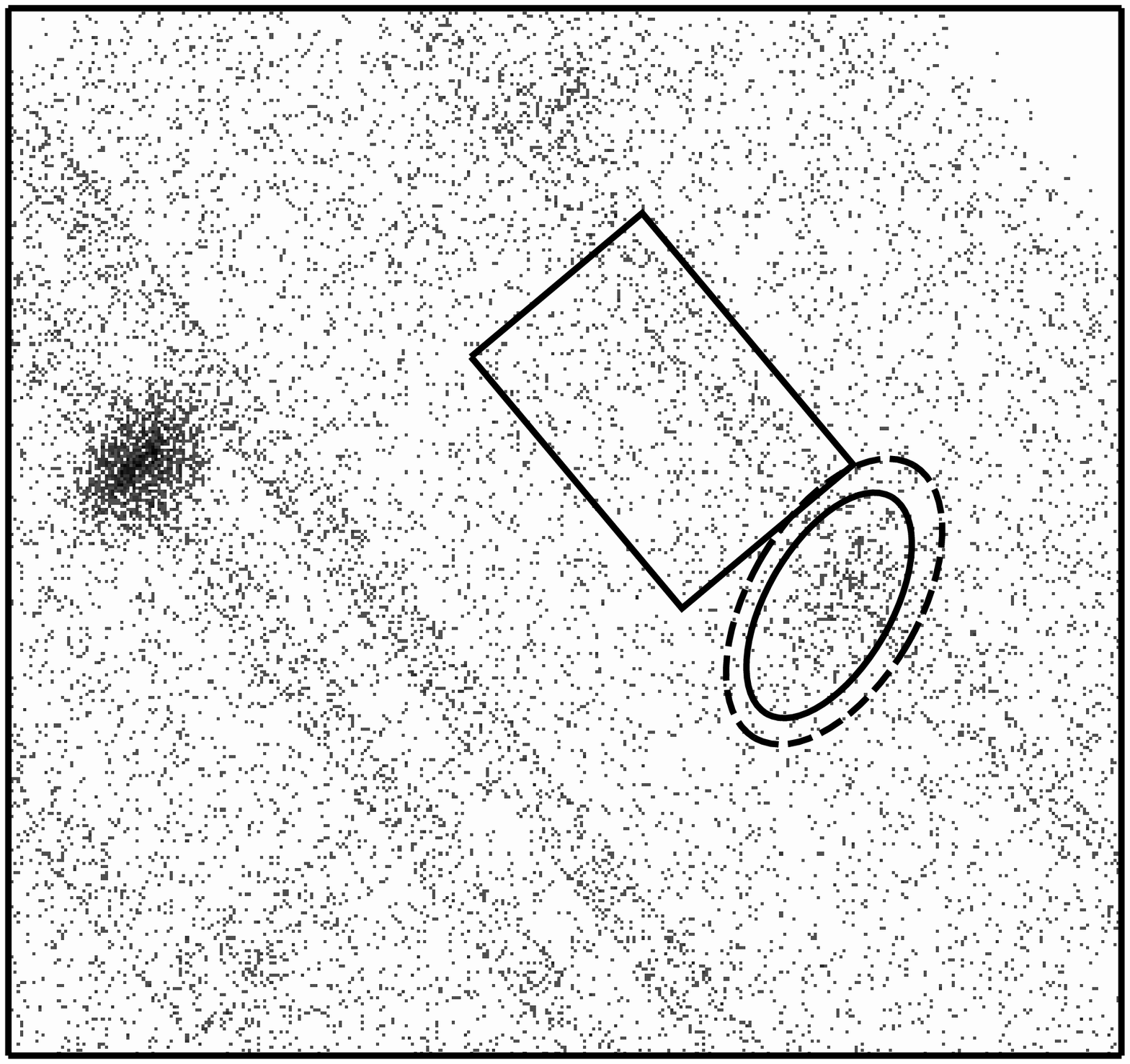} 
\end{minipage}\\[5mm]
\addtocounter{figure}{-1}
\caption{Continued.}
\end{figure*}

\clearpage

\begin{figure}
\includegraphics[height=7cm,bb= 110 40 555 700,angle=-90,clip]{f3.eps}
\caption{\chandra\ ACIS-I spectrum of \xf47\ for the bright state of ObsID 6387. 
The histogram shows the best-fit absorbed power law model.
\label{fig:spec_6387}}
\end{figure}

\begin{thebibliography}{}

\bibitem[Alard 
\& Lupton(1998)]{1998ApJ...503..325A} Alard, C., \& Lupton, R.~H.\ 1998, \apj, 503, 325 

\bibitem[Alard(2000)]{2000A&AS..144..363A} Alard, C.\ 2000, \aaps, 144, 363 

\bibitem[Arnaud(1996)]{1996ASPC..101...17A} Arnaud, K.A.: 1996, {\it 
Astronomical Data Analysis Software and Systems V} 101, 17. 

\bibitem[Bessell et 
al.(1998)]{1998A&A...333..231B} Bessell, M.~S., Castelli, F., \& Plez, B.\ 1998, \aap, 333, 231 

\bibitem[Bildsten et al.(1997)]{1997ApJS..113..367B} Bildsten, L., et al.\ 
1997, \apjs, 113, 367 

\bibitem[Broos et al.(2002)]{Broos2002} Broos, P., Townsley, L., Getman, K., \&
Bauer, F. \ 2002, ACIS Extract, An ACIS Point Source Extraction Package,
Pennsylvania State University

\bibitem[Cardelli et al.(1989)]{1989ApJ...345..245C} Cardelli, J.~A., 
Clayton, G.~C., \& Mathis, J.~S.\ 1989, \apj, 345, 245 

\bibitem[Carter et al. (2003)]{2003ASPC..295..477C}
    Carter, C., Karovska, M., Jerius, D., Glotfelty, K., and Beikman, S.  2003,
    in ASP Conf. Ser. 295: Astronomical Data Analysis  Software and 
    Systems XII, ed.  H. E. Payne, R. I. Jedrzewski, \& R. N. Hook, 477
  
\bibitem[Claret(2004)]{2004A&A...428.1001C} Claret, A.\ 2004, \aap, 428, 1001 

\bibitem[Dickey \& Lockman(1990)]{1990ARA&A..28..215D} Dickey, J.~M., \&
Lockman, F.~J.\ 1990, \araa, 28, 215

\bibitem[Dubus et al.(1999)]{1999MNRAS.302..731D} Dubus, G., Charles, 
P.~A., Long, K.~S., Hakala, P.~J., \& Kuulkers, E.\ 1999, \mnras, 302, 731 
(DCL99) 

\bibitem[Haberl \& Day(1992)]{1992A&A...263..241H} Haberl, F., \& Day, 
C.~S.~R.\ 1992, \aap, 263, 241 

\bibitem[Haberl \& Pietsch(2001)]{2001A&A...373..438H} Haberl, F., \& 
Pietsch, W.\ 2001, \aap, 373, 438 

\bibitem[Hartman et al.(2006)]{2006MNRAS.371.1405H} Hartman, J.~D., 
Bersier, D., Stanek, K.~Z., Beaulieu, J.-P., Kaluzny, J., Marquette, J.-B., 
Stetson, P.~B., \& Schwarzenberg-Czerny, A.\ 2006, \mnras, 371, 1405 

\bibitem[Joye 
\& Mandel(2003)]{2003ASPC..295..489J} Joye, W.~A., \& Mandel, E.\ 2003, Astronomical Data Analysis Software and Systems XII, 295, 489 

\bibitem[Larson \& Schulman(1997)]{1997AJ....113..618L} Larson, D.~T., \& 
Schulman, E.\ 1997, \aj, 113, 618 
 
\bibitem[Long et al.(1981)]{1981ApJ...246L..61L} Long, K.~S., Dodorico, S., 
Charles, P.~A., \& Dopita, M.~A.\ 1981, \apjl, 246, L61 

\bibitem[Makishima et al.(2000)]{2000ApJ...535..632M} Makishima, K., et 
al.\ 2000, \apj, 535, 632 

\bibitem[Markert \& Rallis(1983)]{1983ApJ...275..571M} Markert, T.~H., \& 
Rallis, A.~D.\ 1983, \apj, 275, 571 

\bibitem[Massey et al.(2006)]{2006AJ....131.2478M} Massey, P., Olsen, 
K.~A.~G., Hodge, P.~W., Strong, S.~B., Jacoby, G.~H., Schlingman, W., \& 
Smith, R.~C.\ 2006, \aj, 131, 2478 

\bibitem[Misanovic et al.(2006)]{2006A&A...448.1247M} Misanovic, Z., 
Pietsch, W., Haberl, F., Ehle, M., Hatzidimitriou, D., \& Trinchieri, G.\ 
2006, \aap, 448, 1247 
 
\bibitem[Newton(1980)]{1980MNRAS.190..689N} Newton, K.\ 1980, \mnras, 190,
689

\bibitem[Orosz et al.(2007)]{2007Natur.449..872O} Orosz, J.~A., et al.\ 
2007, \nat, 449, 872 
 
\bibitem[Pietsch et al.(2004b)]{2004A&A...426...11P} Pietsch, W., Misanovic, 
Z., Haberl, F., Hatzidimitriou, D., Ehle, M., \& Trinchieri, G.\ 2004b, 
\aap, 426, 11 (PMH2004)

\bibitem[Pietsch et al.(2004a)]{2004A&A...413..879P} Pietsch, W., Mochejska, 
B.~J., Misanovic, Z., Haberl, F., Ehle, M., \& Trinchieri, G.\ 2004a, \aap, 
413, 879  (PMM2004)


\bibitem[Pietsch et al.(2006b)]{2006ATel..905....1P} Pietsch, W., Plucinsky, 
P.~P., Haberl, F., Shporer, A., \& Mazeh, T.\ 2006b, The Astronomer's 
Telegram, 905, 1 


\bibitem[Pietsch et al.(2006a)]{2006ApJ...646..420P} Pietsch, W., Haberl, 
F., Sasaki, M., Gaetz, T.~J., Plucinsky, P.~P., Ghavamian, P., Long, K.~S., 
\& Pannuti, T.~G.\ 2006a, \apj, 646, 420 


\bibitem[Plucinsky et al.(2008)]{2008ApJS..174..366P} Plucinsky, P.~P., et 
al.\ 2008, \apjs, 174, 366 


\bibitem[Predehl \& Schmitt(1995)]{1995A&A...293..889P} Predehl, P., \&
Schmitt, J.~H.~M.~M.\ 1995, \aap, 293, 889

\bibitem[Pr{\v s}a \& Zwitter(2005)]{2005ApJ...628..426P} Pr{\v s}a, A., \&
Zwitter, T.\ 2005, \apj, 628, 426

\bibitem[Psaltis(2006)]{2006csxs.book....1P} Psaltis, D.\ 2006, in Compact 
stellar X-ray sources (eds. W. Lewin, M. van der Klis, Cambridge University
Press, 1 

\bibitem[Schmidt-Kaler(1982)]{1982Schmidt} Schmidt-Kaler, T.\ 1982, in 
Landolt-B\"ornstein New Series, Vol 2b, Astronomy and astrophysics - Stars 
and star clusters (eds. K. Schaifers, H. H. Voigt), New-York Springer 
Verlag 

\bibitem[Shporer et al.(2007)]{2007A&A...462.1091S} Shporer, A., Hartman, 
J., Mazeh, T., \& Pietsch, W.\ 2007, \aap, 462, 1091 


\bibitem[Shporer 
\& Mazeh(2006)]{2006MNRAS.370.1429S} Shporer, A., \& Mazeh, T.\ 2006, \mnras, 370, 1429 

\bibitem[Shporer et al.(2006)]{2006ATel..913....1S} Shporer, A., Hartman, 
J., Mazeh, T., \& Pietsch, W.\ 2006, The Astronomer's Telegram, 913, 1 


\bibitem[Stark et al.(1992)]{1992ApJS...79...77S} Stark, A.~A., Gammie,
C.~F., Wilson, R.~W., Bally, J., Linke, R.~A., Heiles, C., \& Hurwitz, M.\
1992, \apjs, 79, 77

\bibitem[Stetson(1987)]{1987PASP...99..191S} Stetson, P.~B.\ 1987, \pasp, 
99, 191 

\bibitem[Stetson(1992)]{1992ASPC...25..297S} Stetson, P.~B.\ 1992, in 
ASP Conf. Ser. 25, Astronomical Data Analysis, Software and Systems I, 
eds. D.M. Worrall, C. Biemesderfer, \& J. Barnes (San Francisco: ASP), 297 

\bibitem[van den Bergh(2000)]{2000PASP..112..529V} van den Bergh, S.\ 2000, 
\pasp, 112, 529 

\bibitem[van den Bergh(1991)]{1991PASP..103..609V} van den Bergh, S.\ 1991,
\pasp, 103, 609

\bibitem[Wilms et al.(2000)]{2000ApJ...542..914W} Wilms, J., Allen, A., 
\& McCray, R.\ 2000, \apj, 542, 914 

\bibitem[Wilson \& Devinney(1971)]{1971ApJ...166..605W} Wilson, R.~E., \&
Devinney, E.~J.\ 1971, \apj, 166, 605

\bibitem[Wilson(1979)]{1979ApJ...234.1054W} Wilson, R.~E.\ 1979, \apj, 234,
1054

\bibitem[Wilson(1990)]{1990ApJ...356..613W} Wilson, R.~E.\ 1990, \apj, 356,
613

\bibitem[White et al.(1983)]{1983ApJ...270..711W} White, N.~E., Swank, 
J.~H., \& Holt, S.~S.\ 1983, \apj, 270, 711 

\end{thebibliography}
\end{document}